\documentclass{article} 
\usepackage[preprint]{colm2026_conference}

\usepackage{microtype}
\usepackage{hyperref}
\usepackage{amsmath}
\usepackage[capitalise,noabbrev]{cleveref}
\usepackage{url}
\usepackage{booktabs}
\usepackage{graphicx}
\usepackage{amsmath}
\usepackage{amssymb}

\usepackage{lineno}

\definecolor{darkblue}{rgb}{0, 0, 0.5}
\hypersetup{colorlinks=true, citecolor=darkblue, linkcolor=darkblue, urlcolor=darkblue}

\title{Toward Human-AI Complementarity Across Diverse Tasks}

\author{
Yuzheng Xu \\
The University of Tokyo
\And
Annya Dahmani \\
UC Berkeley
\And
Matthew D. Blanchard \\
University of Sydney
\And
Niclas Dern \\
UC Berkeley
\And
Edy Nastase\textsuperscript{*}
\And
Francesca Bianco\textsuperscript{*}
\And
Maja Pavlovic\textsuperscript{*} \\
Queen Mary University London
\And
Sukanya Krishna\textsuperscript{*} \\
Harvard University
\And
Eric Modesitt\textsuperscript{*} \\
UIUC
\And
Miranda Anna Christ\textsuperscript{*} \\
UC Berkeley
\And
Arth Singh\textsuperscript{*} \\
NIT Agartala
\And
Gaia Molinaro\textsuperscript{*} \\
UC Berkeley
\And
Sikata Bela Sengupta \\
University of Pennsylvania
\And
Jaji Pamarthi
\And
Arjun Menon \\
Princeton University
\And
Rishub Jain\textsuperscript{$\dagger$}
}

\begin{document}

\ifcolmsubmission
\linenumbers
\fi

\maketitle
\renewcommand{\thefootnote}{}
\footnotetext{$^{*}$Equal contribution. $^{\dagger}$Lead and corresponding author. Email: \texttt{rishubjain@gmail.com}. Full author contributions are detailed in Appendix~\ref{app:contributions}.}

\begin{abstract}
Human-AI complementarity, the idea that combining human and AI judgments can outperform either alone, offers a promising pathway toward robust oversight of advanced AI systems.
However, whether human-AI complementarity can be achieved on realistic tasks remains an open question. We investigate this through two approaches: hybridization and two AI assistance methods (top-2 assistance and subtask delegation), evaluated on a multi-domain dataset of 1,886 samples spanning knowledge, factuality, long-context reasoning, and deception detection. We find only modest complementarity gains. Baseline hybridization yields just +0.4 percentage points (pp) over AI alone (69.3\% vs 68.9\%), limited both by a small complementarity region (only 8.9\% of items where AI errs but humans do not) and the inability of confidence-based routing to identify it, since the model's confidence is similarly distributed across correct and incorrect predictions. Applied when AI has low confidence, top-2 assistance increases human accuracy from 28.4\% to 38.3\%, surpassing AI alone (37.7\%) -- but primarily because humans adopt correct AI suggestions, not because they successfully override AI errors. These findings suggest that the primary bottleneck is not human task accuracy per se, but the ability to route decisions to humans when it matters and to design assistance methods that enable humans to catch AI mistakes. Our quantitative and qualitative analyses pinpoint where and why each method succeeds or fails, offering concrete targets for future work. We will release our dataset and code upon request to support progress toward more effective human-AI collaboration for AI oversight.

\end{abstract}

\begin{figure}[t]
\centering
\includegraphics[width=\textwidth]{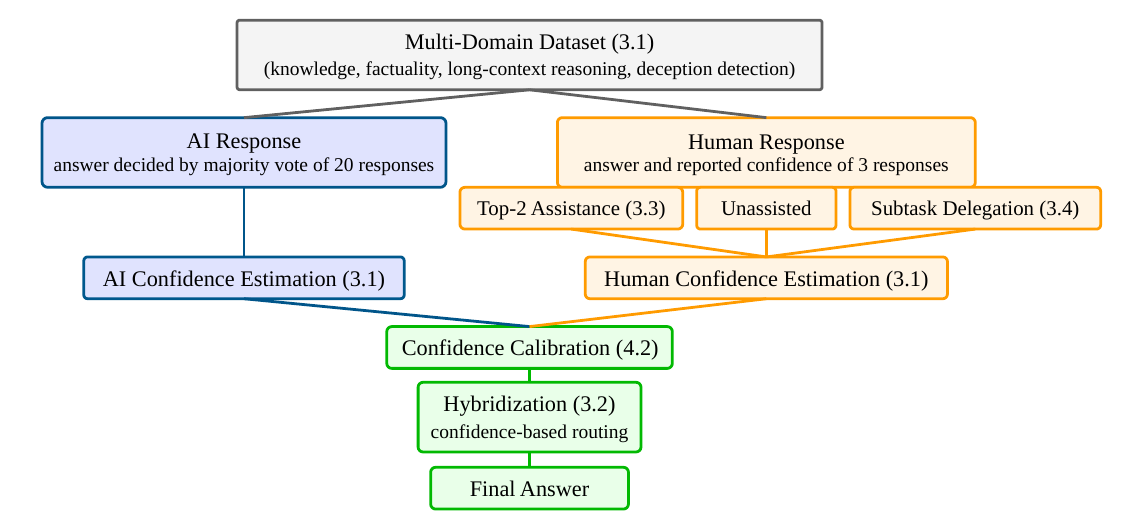}
\caption{\textbf{Overview of our experimental pipeline.} Each item is independently evaluated by AI (majority vote over 20 GPT-5-mini responses \citealp{openai2025gpt5}) and human raters under three conditions: unassisted, with top-2 assistance, and with subtask delegation. Both AI and human confidence scores are calibrated via isotonic regression and used for confidence-based hybridization routing, which selects the AI or human answer per item based on calibrated confidence thresholds. Section numbers in parentheses indicate where each component is described.}
\label{fig:main}
\end{figure}

\section{Introduction}


As AI systems become more capable and are increasingly used for high-stakes tasks, reliably detecting harmful or misaligned behavior becomes a central problem~\citep{bowman2022measuringprogressscalableoversight, leike2018scalableagentalignmentreward}. Oversight -- the monitoring, guidance, and control of AI systems -- matters across training, evaluation, and deployment, since failures at any stage propagate through the pipeline and erode the reliability of downstream systems. 
While AI-based oversight is often more effective and easier to scale, it has clear weaknesses.
AI overseers may share blind spots with the systems they monitor, remain vulnerable to adversarial or deceptive inputs, and be susceptible to reward hacking. They can also exhibit poor calibration, i.e., fail to accurately map confidence to accuracy, in some domains~\citep{engels2025scalinglawsscalableoversight, farquhar2024detecting}. Value judgments about whether a model's behavior aligns with human intentions also require human grounding, particularly in out-of-distribution scenarios where AI cannot reliably determine what humans want.

At the same time, na\"ive human oversight is difficult to scale and often outperformed by AI. 
Human-AI complementarity is based on the idea that humans and AI make \textit{different kinds of errors}, and that selectively combining their judgments can outperform both~\citep{hemmer2024complementarityhumanaicollaborationconcept, steyvers2022_bayesian_complementarity}, offers a promising path forward for robust oversight of advanced AI systems.
The goal of this research agenda is to develop methods for human-AI collaboration that improve the detection of harmful AI behavior now and remain effective as models become more capable. In this work, we test whether this vision of human-AI complementarity holds up across diverse and realistic tasks.

Prior work has shown that complementarity can be achieved in controlled settings and structured interactions, e.g., through explanations or interfaces that reduce overreliance and encourage critical engagement~\citep{bansal2021doesexceedpartseffect, Bu_inca_2021, Kim_2025, vasconcelos2023explanationsreduceoverrelianceai}. To combine human and AI judgments effectively, prior work has developed confidence-based routing methods that learn when to defer to humans or AI~\citep{mozannar2020_learning_to_defer, wilder2020learningcomplementhumans, lee2025_uncertainty_aware_task_delegation} and task decomposition strategies that distribute sub-problems between them~\citep{christiano2018supervisingstronglearnersamplifying, radhakrishnan2023questiondecompositionimprovesfaithfulness}. A recurring challenge is overreliance, where showing humans AI outputs can backfire by discouraging independent judgment~\citep{vasconcelos2023explanationsreduceoverrelianceai, rastogi2022decidingfastslowrole}.

The closest prior result to our work is \citet{jain2025humanaicomplementaritygoalamplified}, who showed that complementarity can be achieved for factuality tasks using hybridization with targeted assistance. A key finding of their work was that simply showing humans model outputs was often ineffective, and that the form of assistance matters. When models are weak, additional, aptly designed support must complement AI suggestions to humans in ways that leverage the human-AI team's complementary strengths. However, \citet{jain2025humanaicomplementaritygoalamplified} studied a single domain and task type. Similarly, much of the broader complementarity literature relies on narrow prediction or classification tasks with relatively simple model scores~\citep{alur2024auditinghumanexpertise, guo2026explainingimprovinginformationcomplementarities}. By contrast, real-world oversight requires detecting harm across many different capabilities, from reading long documents and assessing factuality to applying domain knowledge and catching deception. At present, however, it remains unclear whether complementarity generalizes to all sorts of tasks. Moreover, we have yet to identify the specific tasks where human judgment adds value beyond frontier models, especially given the rise in model capabilities with scale~\citep{engels2025scalinglawsscalableoversight}.

\paragraph{Our Contributions.} To test whether human-AI complementarity generalizes from narrow, single-domain settings to realistic oversight tasks that span multiple domains and capabilities, we build an empirical framework and use it to identify where current approaches succeed, where they fail, and what needs to change.

First, we build a multi-domain evaluation suite of 1,886 samples from 9 sources spanning knowledge, factuality, long-context reasoning, and deception detection tasks, resulting in a highly curated benchmark where a frontier model (GPT-5-mini) outperforms na\"ive humans by roughly 19 percentage points (pp) on average (\cref{sec:methods}). We collect baseline human judgments on our benchmark from over 420 participants, and evaluate them within a confidence-based hybridization framework with calibrated routing (\cref{fig:main}, \cref{sec:experiments}). We find that baseline complementarity gains are small, with hybridization (i.e., simply combining human and AI judgments) achieving +0.4pp over AI alone, constrained by a low complementarity ceiling in which only 8.9\% of items fall in the region where AI is wrong but humans are right.

Building on this baseline, we introduce two AI assistance methods aimed at improving human performance on items where AI is least confident (``low-confidence subset''; \cref{sec:top2,sec:delegation}). In top-2 assistance, humans see the model's two most likely answers along with explanations, while in subtask delegation, questions are decomposed into subtasks, and low-confidence tasks are routed to humans. Both methods improve human accuracy on the low-confidence subset, but for different reasons and with different failure modes. Top-2 assistance boosts accuracy from 28.4\% to 38.3\%, surpassing AI alone at 37.7\%. This gain is concentrated on tasks where AI suggestions are correct (+17.4pp), with no significant improvement when AI is wrong (+5.3pp). 
Subtask delegation helps on tasks that decompose into independent sub-problems (FACTS Search +10pp, QuALITY +25pp, HLE +18pp) but fails on deception detection (+0pp across all three deception datasets), presumably because the decomposition breaks down what the conversation is about rather than whether the AI is acting deceptively. (\cref{sec:qualitative}).

Finally, our qualitative and quantitative analyses (\cref{sec:qualitative}) reveal that human and AI errors are systematically non-overlapping across task types. This leaves room for future improvements but is currently limited by poor calibration. 

Taken together, our results suggest that the primary bottleneck is not human task accuracy per se, but the ability to route decisions to humans when it matters and to design assistance that enables humans to catch AI mistakes. We discuss directions for addressing these challenges and will release our dataset and code upon request to support future research.

\section{Related Work}
\label{sec:related_work}

\paragraph{Human-AI Complementarity and Hybridization.}

Recent work has argued that hybrid systems can outperform either humans or AI in isolation, particularly when their errors are diverse and selectively combined~\citep{hemmer2024complementarityhumanaicollaborationconcept, steyvers2022_bayesian_complementarity}. Complementarity frameworks from HCI and decision-making formalize this intuition by modeling how humans and models contribute different strengths (e.g., pattern recognition at scale versus contextual judgment, value sensitivity, and meta-reasoning)~\citep{hemmer2024complementarityhumanaicollaborationconcept,rastogi2022decidingfastslowrole}. These arguments also hold up empirically, with several studies showing that human-AI teams can outperform either component alone when interaction is structured appropriately, for example when explanations or interfaces reduce overreliance and encourage critical engagement~\citep{bansal2021doesexceedpartseffect,Bu_inca_2021,vasconcelos2023explanationsreduceoverrelianceai,Kim_2025}. A common mechanism for achieving this in practice is hybridization, which selectively routes decisions between humans and AI based on confidence or task characteristics~\citep{bondi_selective_prediction_2022, learning_to_defer_2018, mozannar2020_learning_to_defer, Ibrahim2025}. Various approaches have been proposed, including learning when to defer to human experts~\citep{selective_classification_2017, mozannar2020_learning_to_defer}, directly optimizing joint human-machine team performance~\citep{wilder2020learningcomplementhumans}, delegating tasks based on uncertainty awareness~\citep{lee2025_uncertainty_aware_task_delegation, zhang2024evaluatingutilityconformalprediction}. However, most hybridization methods have been validated on narrow prediction tasks with relatively weak AI systems, leaving their effectiveness in broader settings with highly capable models unclear.

\paragraph{Task Decomposition, Scalable Oversight, and Factored Cognition.}

Task decomposition, which involves breaking down complex problems into simpler subtasks, is a core strategy for scalable oversight. Iterated Distillation and Amplification (IDA)~\citep{christiano2018supervisingstronglearnersamplifying} and Factored Cognition~\citep{sandoval2025factortufactoredcognitionstrengthens} proposed that cognitive reasoning can be algorithmically decomposed into workspaces, where humans answer subtasks, and AI learns to replicate and extend this reasoning, and recursive summarization of books with human feedback~\citep{wu2021recursivelysummarizingbookshuman} demonstrated early success in applying these principles to realistic tasks. Modern decomposition methods vary in execution structure and often trade computational cost against accuracy and reliability~\citep{zhou2023leasttomostpromptingenablescomplex,liu2025selectthendecomposeempiricalanalysisadaptive,hernandezguterrez2025_rdd}. However, not all tasks decompose naturally, and decomposition can introduce new failure modes when subtasks are under-specified or lose context~\citep{khot2023decomposedpromptingmodularapproach,hernandezguterrez2025_rdd}. When decomposition works well, it can improve both the faithfulness of model-generated reasoning~\citep{radhakrishnan2023questiondecompositionimprovesfaithfulness} and human problem-solving. Despite these advances, open questions remain about how decomposition and delegation should be allocated between humans and models in high-capability settings~\citep{tomasev2026_intelligent_ai_delegation}. Recent work has also begun exploring alternative delegation strategies beyond uncertainty or confidence signals, for example by eliciting human preferences over which subtasks should be handled by AI versus humans and training models to imitate these decisions~\citep{feng2026cocoacoplanningcoexecutionai}. Rather than uniformly decomposing all tasks or assigning all subtasks to humans, confidence-guided delegation offers a selective alternative: models handle high-confidence subtasks, while low-confidence components are routed to human oversight~\citep{mozannar2020_learning_to_defer, lee2025_uncertainty_aware_task_delegation}. This framing shifts decomposition from a purely performance-driven strategy to a mechanism for targeted supervision.

\paragraph{Value Alignment and Scalable Oversight.}

A central motivation for human-AI complementarity is its relevance to value alignment in domains where ground truth is contested or undefined. In alignment research, many critical tasks, such as detecting subtle deception, assessing whether model behavior aligns with human values, or evaluating long-term consequences of AI actions, lack clear verifiable solutions~\citep{bowman2022measuringprogressscalableoversight, leike2018scalableagentalignmentreward}. In these settings, human judgment remains important, but humans may struggle with complex or adversarial model outputs. Scalable oversight frameworks address this challenge by designing systems where less capable humans can meaningfully supervise more capable AI~\citep{bowman2022measuringprogressscalableoversight, christiano2018supervisingstronglearnersamplifying}, with recursive reward modeling~\citep{leike2018scalableagentalignmentreward} and debate-based oversight~\citep{irving2018aisafetydebate} as prominent instantiations, though empirical validations remain limited. As AI systems approach or surpass human raters on many evaluation tasks, a central question is whether hybrid human-AI systems can maintain meaningful oversight when AI alone outperforms people. This goes beyond static accuracy on individual tasks. Alignment-relevant complementarity also concerns the robustness of the oversight signal under distributional change and optimization pressure, and identifying the regimes in which human input improves reliability remains an open empirical problem~\citep{engels2025scalinglawsscalableoversight}.

\section{Methods}
\label{sec:methods}

Our pipeline is shown in \cref{fig:main}. We first build a comprehensive benchmark from independent sources. Then, we present each item to AI alone (GPT-5-mini) and human raters under three conditions: unassisted, with top-2 assistance, and with subtask delegation. Both AI and human confidence scores are calibrated and used for confidence-based hybridization routing, which decides whether to use the AI or human answer for each item. We now describe each component in detail.

\subsection{Dataset and Task Setup}

We constructed a multi-domain dataset spanning knowledge, factuality, long-context reasoning, and deception detection by selecting items from existing benchmarks (see \cref{app:datasets} for selection criteria and preparation details). The final dataset comprises 1,886 items from 9 sources: QuALITY~\citep{pang-etal-2022-quality} hard subset (300), BIG-Bench~\citep{srivastava2023beyond} (387), Humanity’s Last Exam~\citep{phan_benchmark_2026} (147), GPQA Diamond~\citep{rein2024gpqa} (146), SimpleQA Verified~\citep{haas2025simpleqa} (300), FACTS Search~\citep{cheng2025facts} (300), Hidden Agenda~\citep{hiddenagenda} (100), SHADE-Arena~\citep{kutasov2025shadearenaevaluatingsabotagemonitoring} (106), and Web of Lies~\citep{Makovi2021} (100). Items are formatted as either multiple-choice (MC) or free-text (FT) questions. SimpleQA was excluded from hybridization experiments due to near-ceiling AI accuracy (98.3\%), leaving 1,586 items for hybridization analysis.

For each item, we collected both AI responses and human judgments. The AI model (GPT-5-mini) generated 20 independent responses per item, each including a self-reported confidence score. Human participants were recruited via Prolific~\citep{Prolific2025} and answered questions in Gorilla~\citep{Gorilla2026}, providing both an answer and a confidence rating. We collected 3-5 independent human responses per question. See \cref{app:exp_design} for details about human data collection.

\paragraph{AI Confidence Estimation.}
To obtain a single AI answer and confidence per item, we aggregate over the 20 sampled GPT-5-mini responses. For MC questions, the final answer is the most frequent option (ties broken by mean confidence). For FT questions, we group semantically equivalent responses using the semantic entropy method~\citep{farquhar2024detecting}, and select the largest cluster (ties broken by mean confidence). We then compute confidence in two ways:
\begin{itemize}
    \item \textbf{Direct-ask confidence}: The mean self-reported confidence (0--1) across responses that match the majority answer (MC) or belong to the winning cluster (FT).
    \item \textbf{Computed confidence}: For MC questions, the fraction of responses agreeing with the majority answer (self-consistency). For FT questions, $1 - H_{\text{norm}}$, where $H_{\text{norm}}$ is the normalized Shannon entropy over semantic clusters, following~\citet{farquhar2024detecting}.
\end{itemize}
We compared both methods using calibration metrics on a held-out split and selected direct-ask confidence with isotonic regression calibration for all downstream experiments, as it achieved the lowest Brier score (0.174; see \cref{sec:experiments,app:calibration} for details).

\paragraph{Human Confidence.}
We analyze human confidence at two levels. In the majority-vote setting, we take the mean confidence of participants whose answer matches the majority answer for each item, then calibrate this score using isotonic regression. In the individual setting, each participant’s raw confidence is calibrated independently using the same isotonic regression model. Since some items received up to 5 responses while others received 3, we limit each item to at most 3 responses (selected deterministically by participant ID, setting the random seed to 42) so that no item is overrepresented in the analysis.

\paragraph{Data Selection.}
We use two subsets of the data for evaluation. The full set includes all 1,586 items (excluding SimpleQA), where baseline human data is available for every item. The low-confidence subset targets items where AI is most likely to benefit from human input. We selected items with the lowest calibrated AI confidence per dataset (bottom 10\% per dataset, 20\% for Big-Bench), yielding a 191-item ``low-confidence subset'' after excluding 4 questions with incomplete data. Human data for the top-2 assistance and subtask delegation conditions were collected only on this subset. For each subset, we use a 40\% split for fitting calibration models and selecting routing thresholds and a 60\% test split on which we report our results. For the low-confidence subset, this yields 69 calibration and 122 test items.\footnote{Post hoc, we found inconsistent tie-breaking between answer selection and confidence, affecting 14 items. As fixing this would change the subset and require new data, we report the original results and defer correction to the next version.}

\subsection{Hybridization}
\label{sec:hybridization}

All hybridization methods are evaluated against AI-alone and human-alone baselines. Both AI and human confidence are calibrated via isotonic regression models trained on the 40\% calibration set (634 items). We test three routing methods, each of which uses confidence thresholds optimized per dataset group on the calibration set to maximize hybrid accuracy.

\begin{itemize}
    \item \textbf{1-threshold}: Use AI if $\text{AI\_conf} \geq T$, otherwise defer to human. $T$ is selected by grid search (step=0.01, range $[0, 1]$).
    \item \textbf{2-threshold}: Use human if $\text{Human\_conf} > T_h$ AND $\text{AI\_conf} < T_a$, otherwise use AI. Both thresholds are selected by grid search ($101 \times 101$ combinations).
    \item \textbf{1-threshold + compare}: Use AI if $\text{AI\_conf} \geq T$ \emph{or} $\text{AI\_conf} > \text{Human\_conf}$, otherwise defer to human. $T$ is selected identically to the 1-threshold method. This combines confidence thresholding with a direct confidence comparison, routing to humans only when both signals agree that the AI is uncertain.
    \item \textbf{Compare}: Use AI if $\text{AI\_conf} > \text{Human\_conf}$, otherwise defer to human. This threshold-free method routes decisions purely based on which source has higher calibrated confidence.
    
\end{itemize}

For individual-level analysis, each participant’s calibrated confidence is used independently for the hybrid routing decision, so different participants on the same item may be routed differently. The three deception datasets (Hidden Agenda, SHADE-Arena, Web of Lies) are combined into a single group for threshold learning due to small individual calibration set sizes.

\subsection{Top-2 Assistance}
\label{sec:top2}

In the top-2 assistance condition, participants receive AI-generated guidance before answering. For each item, GPT-5-mini generates its two most likely answers, each accompanied by a brief explanation. Participants see both candidate answers alongside the original question and then provide their own answer and confidence rating. The goal is to test whether exposure to AI reasoning helps humans make better judgments, particularly on items where AI is uncertain.

Human data for this condition was collected on the same 191-item low-confidence subset described above. On the full test set, items outside this subset use the AI answer directly, since AI confidence on those items exceeds any routing threshold. Hybridization within the subset follows the same three routing methods, with thresholds selected on the subset’s calibration portion.

\subsection{Subtask Delegation}
\label{sec:delegation}

In the subtask delegation condition, each item is broken into smaller sub-problems that can be handled independently by AI or humans. GPT-5-mini decomposes each item into several subtasks, answers each one, and reports a confidence score (0--1) per subtask. Subtasks where model confidence falls below 0.8 are flagged for human input. Participants see the full decomposition with the AI’s proposed answers and provide their own answers for low-confidence subtasks (they may also optionally override high-confidence ones). After all subtask answers are collected from both humans and AI, GPT-5-mini synthesizes them into a final answer via a recomposition call. The goal is to concentrate human effort on the specific sub-problems where AI is least certain, rather than requiring humans to answer the full question. Human data was collected on the same 191-item low-confidence subset as top-2 assistance.

\section{Experimental Results}
\label{sec:experiments}

We first report baseline human and AI performance across all datasets, then present hybridization and assistance results on both the full test set (952 items) and the 122-item low-confidence test subset, where all three human conditions are directly comparable.

\subsection{Baseline Performance}

Before hybridization, we evaluate how humans and AI perform independently. \Cref{tab:baseline} summarizes accuracy across all 9 datasets on the test set.

\begin{table}[t]
\centering
\begin{tabular}{llrrrr}
\toprule
Dataset & Type & N (test) & Human Maj & Human Ind & AI Acc \\
\midrule
FACTS Search & FT & 180 & 69.4\% & 58.5\% & 86.7\% \\
QuALITY & MC & 180 & 61.1\% & 51.5\% & 80.6\% \\
Big-Bench & MC/FT & 232 & 29.7\% & 24.0\% & 61.6\% \\
GPQA Diamond & MC & 88 & 44.3\% & 38.3\% & 68.2\% \\
Hidden Agenda & MC & 60 & 83.3\% & 77.8\% & 81.7\% \\
Humanity's Last Exam & MC/FT & 88 & 12.5\% & 9.5\% & 19.3\% \\
SHADE-Arena & MC & 64 & 42.2\% & 43.8\% & 57.8\% \\
Web of Lies & MC & 60 & 73.3\% & 73.3\% & 81.7\% \\
SimpleQA Verified* & FT & 180 & 81.7\% & 79.8\% & 98.3\% \\
\midrule
\textbf{Overall (excl.\ SimpleQA)} & -- & \textbf{952} & \textbf{49.9\%} & \textbf{43.5\%} & \textbf{68.9\%} \\
\bottomrule
\end{tabular}
\caption{\textbf{Baseline accuracy comparison.} Human and AI (GPT-5-mini) accuracy on the test set (N=952 excluding SimpleQA). SimpleQA was excluded from hybridization experiments due to high AI accuracy (98.3\%), leaving insufficient headroom for improvement.}
\label{tab:baseline}

\end{table}

AI consistently outperforms humans across most datasets, with an average gap of +19.0 percentage points. The largest gaps appear on Big-Bench (+31.9\%) and GPQA Diamond (+23.9\%), while Hidden Agenda is the only dataset where humans slightly outperform AI ($-$1.7\%). Notably, Humanity's Last Exam is challenging for both humans and AI (below 20\% accuracy), suggesting fundamental task difficulty rather than differential capability. In the individual setting, where each participant's response is evaluated separately rather than aggregated by majority vote, human accuracy is generally lower, except for SHADE-Arena. Aggregating multiple human responses via majority vote consistently improves accuracy.

\paragraph{Results on the Low-Confidence Subset.}
Human data for top-2 assistance and subtask delegation was collected only on the 191-item low-confidence subset. After applying the 40/60 calibration-test split, 122 items remain in the test portion with all three human conditions (baseline, top-2, delegation) available. 
In addition to the three threshold-based routing methods, we also evaluate the compare method (\cref{sec:hybridization}), which routes purely on relative calibrated confidence. All three hybrid conditions outperform both human alone and AI alone on this subset. 
Logistic mixed-effects models on individual-level data confirm that all three are significantly more accurate than human alone and AI alone (all $p$s $< .001$), but do not differ significantly from one another (\cref{tab:mixed_model,tab:pairwise}). However, none of the effects reached significance in the majority-vote model.
Top-2 assistance produces the highest raw human accuracy (38.3\%), surpassing AI alone (37.7\%). The compare method achieves better results for baseline and top-2 conditions on majority voting, but does not yield complementarity when combined with subtask delegation. \cref{tab:subset_perdataset} summarizes the accuracy of each condition on the 122-item test subset, and \cref{tab:oracle} shows the oracle routing results. For majority voting, the oracle ceiling is 53.3\% (from the baseline method). For individual responses, the oracle ceiling is from top-2 assistance.

\subsection{Hybridization and Assistance Results}

\paragraph{Confidence Calibration.}
To convert raw confidence into calibrated probabilities, we trained isotonic regression and Platt scaling models on the 40\% calibration set (see \cref{app:calibration} for details). Based on the Brier score, we selected direct-ask confidence with isotonic regression as the confidence source for downstream hybridization. All reported accuracies are evaluated exclusively on the held-out 60\% test set.

\paragraph{Full-Set Results.}
\cref{tab:hybrid_main} summarizes hybridization results across all three conditions on the full test set. For top-2 assistance and subtask delegation, items outside the low-confidence subset default to AI.

Baseline 2-threshold and baseline 1-threshold+compare both achieve the best hybrid accuracy (69.3\% majority), a +0.4pp improvement over AI alone. For top-2 assistance, the 1-threshold+compare method matches this at 69.3\% majority, compared to 69.1\% for 1-threshold alone. The 1-threshold+compare method routes to humans only when both the threshold and compare signals agree that the AI is uncertain, producing a more selective routing policy. All hybrid conditions show comparable gains (+0.1--0.4pp). The gains are modest because ${\sim}$93--99\% of items use AI across all conditions.

\paragraph{Why Only Limited Hybridization Gains?}

The modest hybridization gains (+0.2--0.4pp) can be attributed to several factors. First, AI outperforms humans by ${\sim}$20pp on average. When AI is substantially more accurate, deferring to human judgment is rarely beneficial. Second, while calibration improves the reliability of confidence estimates, the correlation between calibrated AI confidence and accuracy is not strong enough to reliably identify the subset of examples where humans would outperform AI.
On the deception datasets specifically, all confidence methods collapse to AUROC $\approx$ 0.50, meaning confidence carries no discriminative signal for these safety-relevant tasks (\cref{app:confidence}).
Third, for hybridization to yield large gains, humans and AI should make errors on different examples. In our data, many examples that AI gets wrong are also difficult for humans, limiting the complementarity opportunity. The exceptions are Hidden Agenda and SHADE-Arena, where human and AI errors are more complementary (see \cref{app:disagreements}).

\begin{table}[t]
\centering
\begin{tabular}{lrrrrr}
\toprule
 & Human & AI & Baseline (best) & Top-2 (best) & Subtask (best) \\
\midrule
\textbf{Maj.} & 49.9\% & 68.9\% & 69.3\% (2T/1TC) & 69.3\% (1TC) & 69.2\% (1T) \\
\textbf{vs AI} & $-$19.0pp & --- & +0.4pp & +0.4pp & +0.3pp \\
\textbf{Ind.} & 43.5\% & 68.9\% & 69.2\% (2T) & 69.1\% (1T/1TC) & 69.2\% (1T) \\
\textbf{vs AI} & $-$25.4pp & --- & +0.3pp & +0.2pp & +0.3pp \\
\bottomrule
\end{tabular}
\caption{\textbf{Hybridization results on the full test set (N=952).} Best routing method per condition, selected from 1-threshold (1T), 2-threshold (2T), and 1-threshold+compare (1TC).}
\label{tab:hybrid_main}
\end{table}

We also find evidence of overreliance when humans are shown AI outputs directly. In the low-confidence subset (where human input matters the most), top-2 assistance significantly improved human accuracy on items the AI answered correctly ($+$17.4\,pp, $p = 0.006$) but produced no significant gain on items the AI answered incorrectly ($+$5.3\,pp, $p = 0.213$; see \cref{fig:overreliance-triangles}). Subtask delegation, where participants answer decomposed subtasks without viewing model answers, did not exhibit this asymmetry. This pattern of results suggests that direct exposure to model outputs induces overreliance that limits the value of human input precisely where hybridization needs it most (see \cref{app:overreliance_quantitative} for the full analysis and \cref{app:top2_qualitative} for qualitative examples).

\section{Qualitative Analyses}
\label{sec:qualitative}
To understand where and why our methods succeed or fail, we conducted three qualitative analyses. First, we examined human-AI disagreement patterns across datasets to identify where human and AI strengths differ systematically. Second, we analyzed when top-2 assistance helps or hinders human judgment. Third, we investigated where subtask delegation succeeds and where it breaks down. We summarize our findings below, with full details in \cref{app:disagreements,app:top2_qualitative,app:subtask_qualitative}.

In three datasets, human and AI errors cluster around distinct capability profiles. In FACTS Search, humans succeed on simple, directly searchable single-hop questions while AI handles multi-hop reasoning well. In Hidden Agenda, humans catch social engineering patterns while AI detects technical threats. In SHADE-Arena, humans spot intuitive red flags from everyday fraud patterns while AI catches sabotage embedded in routine operations. These patterns suggest that human and AI errors are not random but reflect different underlying capabilities, which a better routing mechanism could exploit.

Top-2 assistance helps humans most when AI suggestions reduce verification effort, for example, by narrowing a long text to a specific hypothesis or providing a near-miss answer that is easy to correct. However, when AI suggestions are incorrect, they can act as distractors that undermine careful reasoning, particularly on tedious tasks where participants use the suggestion as a shortcut rather than verifying it independently. We also find that participants anchor on the first AI suggestion and discount the second, even when the second is correct.

Subtask delegation shows localized gains on tasks that decompose into independent sub-problems, including FACTS Search (+10pp), QuALITY (+25pp), and HLE (+18pp), but fails on deception detection (0\% across all three deception datasets) and GPQA Diamond ($-$41.7pp). The failure on deception tasks occurs because the model decomposes the surface content of conversations (e.g., gardening advice subtasks) rather than the evaluation objective (whether the AI is behaving suspiciously), so the detection question is never surfaced as a subtask. On average, each item was decomposed into 8.86 subtasks, of which 20\% were routed to humans, with routing rates varying from 0.5\% on QuALITY to 48.6\% on FACTS Search. 

\section{Conclusion}
\label{sec:future}

In this work, we set out to evaluate whether human-AI complementarity can be achieved in realistic settings where models already perform strongly alone. To test this, we constructed a multi-domain benchmark spanning domain knowledge, factuality, long-context reasoning, and deception detection, where AI substantially outperforms human raters by roughly 19 percentage points on average. Based on this benchmark, we developed confidence-based hybridization strategies and two AI assistance methods (top-2 assistance and subtask delegation), and collected ratings from over 420 human participants to evaluate them. The best hybridization approach achieved 69.3\% accuracy (+0.4pp over AI alone), with top-2 assistance at +0.2pp and subtask delegation at +0.2pp.

While these gains are modest, our analyses suggest concrete paths forward for achieving complementarity. Qualitative analyses reveal that human and AI errors are systematically non-overlapping across task types, with deception detection showing the sharpest divergence and thus being a particularly promising domain for future work. A key barrier to exploiting these error patterns is overreliance: top-2 assistance significantly improved human accuracy when AI was correct, but provided no meaningful benefit when AI was wrong, suggesting that future assistance methods must focus on helping humans recognize when to override AI instead of blindly deferring to it.

Several limitations should be noted. The study was underpowered to detect the observed +0.4pp effect, with post-hoc power estimated at 5.4\%. Additionally, our benchmark relies heavily on reused public datasets, raising the possibility that GPT-5-mini has been exposed to these items during training. Furthermore, all human participants were lay raters recruited via Prolific, not domain experts, and the complementarity picture could look very different with expert participants, particularly on tasks like GPQA Diamond and Humanity's Last Exam where domain knowledge is essential. Similarly, all results are based on a single AI model (GPT-5-mini), and different models may exhibit different error patterns and calibration properties. Finally, we tested only one form of top-2 assistance and one decomposition strategy, while the design space of possible assistance interfaces is much larger.

These limitations also point to natural directions for future work. Beyond exploring expert participants, multiple models, and richer assistance interfaces, future work should focus on routing signals beyond confidence (such as activation-based probes that detect when models are confidently wrong), task-type-aware decomposition prompts that preserve the evaluation objective for deception detection tasks, and assistance designs that help humans recognize when to override AI rather than simply showing them model outputs. We will release all data and code upon request to support future research on human-AI complementarity and scalable oversight; interested researchers should contact the corresponding author.

\section*{Acknowledgments}
This research was conducted as part of \href{https://sparai.org}{SPAR Fall 2025}, an AI safety research program. They provided funding and recruiting infrastructure, and the project would not have existed without them.

We thank Jess Bergs for being involved in early discussions and helping understand external infrastructure, and William Overman for being involved in early discussions and helping identify datasets.

\section*{Disclosure of LLM Use}
LLMs were used to assist in writing, grammar correction, and restructuring portions of the text. They also helped generate some content (subsequently edited by the authors), create LaTeX tables, generate code for result tables, and verify table outputs. LLMs were additionally used as a coding assistant. LLMs were also used to review the draft to help improve the overall structure of the paper. This disclosure statement was also written with LLM assistance.

\section*{Ethics Statement}
Participation was entirely voluntary, and participants provided informed consent before beginning the task. They were informed of their right to withdraw at any point, and that their responses would remain anonymous. No personally identifiable information was collected and Prolific IDs were used only for compensation. There were no known risks in our study and participants were compensated fairly for their time.

\newpage
\bibliography{colm2026_conference}
\bibliographystyle{colm2026_conference}

\newpage
\appendix
\crefalias{section}{appendix}

\section{Author Contributions}
\label{app:contributions}
\textbf{Yuzheng Xu}: Built confidence calibration and hybridization methods; developed delegation assistant MVP; managed datasets, visualization, and co-coordinated paper writing/submission. \\
\textbf{Annya Dahmani}: Contributed to dataset curation, methodology, analyses, and project administration, and served as a lead writer on the original draft and review and editing of the final paper. \\
\textbf{Matthew D. Blanchard}: Led end-to-end human data collection: power analysis, participant recruitment, Gorilla study design, data processing, and statistical analysis; also contributed to dataset selection and methods and results writing. \\
\textbf{Niclas Dern}: Contributed to methodology and data curation, co-led development of the Gorilla study interface, and built the subtask delegation interaction interface. Contributed to writing the final paper. \\
\textbf{Edy Nastase}: Implemented subtask delegation method, aligning with hybridisation and UI requirements. Conducted Factored Cognition literature review. \\
\textbf{Francesca Bianco}: Contributed to dataset scouting and review; owned dataset processing and final dataset-merging pipelines; supported literature synthesis and methodology design; contributed to Gorilla integration; and ran confidence calibration experiments. \\
\textbf{Maja Pavlovic}: Led the development of the top-2 assistance method; co-led the design of the gorilla study interface including end-to-end testing. \\
\textbf{Sukanya Krishna}: Contributed to data curation, engineering, and formatting; contributed to hybridization and logging code; and led writing of related work. \\
\textbf{Eric Modesitt}: Ran AI confidence estimation, showing 39 methods' calibration and discrimination properties. Designed the 1-threshold + compare routing method and contributed results, tables, and statistical analyses to the paper. \\
\textbf{Miranda Anna Christ}: Led the quantitative and qualitative analysis of human-AI disagreements for the baseline data, and conducted some initial AI confidence analysis. Designed and produced the paper's main figure. \\
\textbf{Arth Singh}: Led the analysis for quantitative overreliance, GPQA positional bias, and oracle hybridization bounds. \\
\textbf{Gaia Molinaro}: Performed human confidence calibration analyses, qualitative analyses on the top-2 assistance method, and helped with writing and editing. \\
\textbf{Sikata Bela Sengupta}: Implemented dataset scaffolding, review of human instructions, proposals/lit reviews of adaptive hybridization techniques, confidence-based human-ai teaming methods, rl/contextual bandits, analysis of hard examples, and paper structuring. \\
\textbf{Jaji Pamarthi}: Contributed to dataset preparation, review, and study setup; contributed methods and metrics review. \\
\textbf{Arjun Menon}: Conducted literature review and sourced datasets for selection. Processed input data and implemented evaluation code for running hybridization methods. \\
\textbf{Rishub Jain}: Conceived, recruited for, and directed the project across all workstreams.

\section{Confidence Calibration}
\label{app:calibration}
We evaluated two AI confidence estimation methods: direct-ask and computed confidence. Both methods were calibrated using Platt Scaling and Isotonic Regression on held-out calibration sets.

\begin{table}[h]
\centering
\begin{tabular}{llrrr}
\toprule
Confidence Method & Calibration & ECE $\downarrow$ & Brier $\downarrow$ & Accuracy \\
\midrule
ai\_direct & Original & 0.182 & 0.211 & 68.9\% \\
ai\_direct & Platt & 0.048 & 0.179 & 68.9\% \\
ai\_direct & Isotonic & 0.077 & 0.174 & 68.9\% \\
ai\_computed (MC) & Original & 0.182 & 0.218 & 71.0\% \\
ai\_computed (MC) & Platt & 0.065 & 0.190 & 71.0\% \\
ai\_computed (MC) & Isotonic & 0.080 & 0.191 & 71.0\% \\
human & Original & 0.158 & 0.253 & 49.9\% \\
human & Platt & 0.052 & 0.223 & 49.9\% \\
human & Isotonic & 0.038 & 0.222 & 49.9\% \\
\bottomrule
\end{tabular}
\caption{Calibration metrics for AI confidence methods (40/60 split, evaluated on test set). ECE = Expected Calibration Error, Brier = Brier Score. Lower is better for both metrics.}
\label{tab:calibration}
\end{table}

\cref{tab:calibration} shows the calibration results. For ai\_direct confidence, Platt scaling achieves the lowest ECE (0.048), indicating better calibration, while Isotonic regression achieves the lowest Brier score (0.174), indicating better overall prediction quality. We selected Isotonic regression for our downstream tasks because the Brier score captures both calibration and sharpness, which is important for identifying low-confidence samples for two AI assistance tasks.

\subsection{Per-Dataset Results}
\begin{table}[h]
\centering
\resizebox{\textwidth}{!}{%
\begin{tabular}{lrrrrrrrrrrrr}
\toprule
Dataset & N & Human & AI & Base 1T & Base 2T & Base 1TC & Top2 1T & Top2 2T & Top2 1TC & Sub 1T & Sub 2T & Sub 1TC \\
\midrule
\multicolumn{13}{l}{\textbf{Majority vote}} \\
\midrule
FACTS & 180 & 69.4\% & 86.7\% & 88.3\% & 88.3\% & 88.3\% & 86.7\% & 86.7\% & 87.2\% & 87.2\% & 87.2\% & 87.2\% \\
QuALITY & 180 & 61.1\% & 80.6\% & 80.6\% & 80.6\% & 80.6\% & 80.6\% & 80.6\% & 80.6\% & 80.6\% & 80.6\% & 80.6\% \\
bbeh & 232 & 29.7\% & 61.6\% & 61.6\% & 61.6\% & 61.6\% & 61.6\% & 61.2\% & 62.1\% & 62.9\% & 62.9\% & 62.5\% \\
gpqa & 88 & 44.3\% & 68.2\% & 68.2\% & 69.3\% & 69.3\% & 70.5\% & 70.5\% & 70.5\% & 65.9\% & 65.9\% & 64.8\% \\
Hidden Agenda & 60 & 83.3\% & 81.7\% & 81.7\% & 81.7\% & 81.7\% & 81.7\% & 83.3\% & 81.7\% & 81.7\% & 81.7\% & 81.7\% \\
SHADE-Arena & 64 & 42.2\% & 57.8\% & 57.8\% & 57.8\% & 57.8\% & 57.8\% & 56.2\% & 57.8\% & 57.8\% & 57.8\% & 57.8\% \\
Web of Lies & 60 & 73.3\% & 81.7\% & 81.7\% & 81.7\% & 81.7\% & 81.7\% & 83.3\% & 81.7\% & 81.7\% & 81.7\% & 81.7\% \\
HLE & 88 & 12.5\% & 19.3\% & 19.3\% & 19.3\% & 19.3\% & 19.3\% & 19.3\% & 19.3\% & 19.3\% & 19.3\% & 20.5\% \\
\midrule
\textbf{All} & \textbf{952} & \textbf{49.9\%} & \textbf{68.9\%} & \textbf{69.2\%} & \textbf{69.3\%} & \textbf{69.3\%} & \textbf{69.1\%} & \textbf{69.1\%} & \textbf{69.3\%} & \textbf{69.1\%} & \textbf{69.1\%} & \textbf{69.0\%} \\
\midrule
\multicolumn{13}{l}{\textbf{Individual responses}} \\
\midrule
FACTS & 540 & 58.5\% & 86.7\% & 87.6\% & 87.8\% & 87.2\% & 87.0\% & 87.0\% & 87.0\% & 87.8\% & 86.8\% & 87.4\% \\
QuALITY & 540 & 51.5\% & 80.6\% & 80.6\% & 80.6\% & 80.6\% & 80.6\% & 80.6\% & 80.6\% & 80.9\% & 80.9\% & 80.6\% \\
bbeh & 696 & 24.0\% & 61.6\% & 61.5\% & 61.8\% & 61.5\% & 61.5\% & 61.6\% & 61.5\% & 62.2\% & 62.2\% & 62.1\% \\
gpqa & 264 & 38.3\% & 68.2\% & 68.9\% & 68.2\% & 67.0\% & 69.7\% & 69.3\% & 69.7\% & 66.7\% & 66.7\% & 65.9\% \\
Hidden Agenda & 180 & 77.8\% & 81.7\% & 81.7\% & 81.7\% & 81.7\% & 81.7\% & 81.7\% & 81.7\% & 81.7\% & 81.7\% & 81.7\% \\
SHADE-Arena & 192 & 43.8\% & 57.8\% & 57.8\% & 57.8\% & 57.8\% & 57.8\% & 56.2\% & 57.8\% & 57.8\% & 57.8\% & 57.8\% \\
Web of Lies & 180 & 73.3\% & 81.7\% & 81.7\% & 81.7\% & 81.7\% & 81.7\% & 81.7\% & 81.7\% & 81.7\% & 81.7\% & 81.7\% \\
HLE & 264 & 9.5\% & 19.3\% & 19.3\% & 19.3\% & 19.3\% & 19.3\% & 19.3\% & 19.3\% & 19.3\% & 19.3\% & 20.1\% \\
\midrule
\textbf{All} & \textbf{2856} & \textbf{43.5\%} & \textbf{68.9\%} & \textbf{69.1\%} & \textbf{69.2\%} & \textbf{68.8\%} & \textbf{69.1\%} & \textbf{69.0\%} & \textbf{69.1\%} & \textbf{69.2\%} & \textbf{69.0\%} & \textbf{69.0\%} \\
\bottomrule
\end{tabular}%
}
\caption{Per-dataset hybridization results (40/60 split). Top panel: majority vote (N=952 items). Bottom panel: individual responses (N=2856 observations). 1TC = 1-threshold + compare.}
\label{tab:hybrid_perdataset}

\end{table}

\cref{tab:hybrid_perdataset} shows the per-dataset breakdown for majority vote and individual responses. The three deception datasets are shown separately to highlight their different characteristics: Hidden Agenda is the only dataset where humans outperform AI in majority vote (83.3\% vs 81.7\%), while SHADE-Arena has the lowest human accuracy (42.2\% majority, 43.8\% individual). Hybridization gains are concentrated in datasets where the AI-human accuracy gap is smaller and human accuracy is non-trivial: FACTS (+1.6pp baseline majority), gpqa (+2.3pp top-2 majority), and bbeh (+1.3pp subtask majority). Datasets where AI is dominant (FACTS, QuALITY) or both perform poorly (HLE) show minimal change.

\subsection{Results on the Low-Confidence Subset}
\begin{table}[h]
\centering
\resizebox{\textwidth}{!}{%
\begin{tabular}{lrrrrrrrrrrrrrrr}
\toprule
Dataset & N & Human & T2 Raw & AI$^\dagger$ & B 1T & B 2T & B 1TC & T2 1T & T2 2T & T2 1TC & S 1T & S 2T & S 1TC \\
\midrule
\multicolumn{14}{l}{\textbf{Majority vote}} \\
\midrule
FACTS & 20 & 45.0\% & 30.0\% & 30.0\% & 45.0\% & 45.0\% & 45.0\% & 30.0\% & 30.0\% & 35.0\% & 35.0\% & 35.0\% & 36.8\% \\
QuALITY & 16 & 37.5\% & 56.2\% & 62.5\% & 62.5\% & 62.5\% & 62.5\% & 62.5\% & 62.5\% & 62.5\% & 62.5\% & 62.5\% & 71.4\% \\
bbeh & 46 & 28.3\% & 23.9\% & 26.1\% & 26.1\% & 26.1\% & 26.1\% & 26.1\% & 23.9\% & 28.3\% & 32.6\% & 32.6\% & 30.2\% \\
gpqa & 12 & 50.0\% & 50.0\% & 33.3\% & 33.3\% & 33.3\% & 50.0\% & 50.0\% & 50.0\% & 50.0\% & 16.7\% & 16.7\% & 8.3\% \\
Hidden Agenda & 7 & 57.1\% & 100\% & 85.7\% & 85.7\% & 57.1\% & 85.7\% & 85.7\% & 100\% & 85.7\% & 85.7\% & 85.7\% & 85.7\% \\
HLE & 11 & 0.0\% & 18.2\% & 0.0\% & 0.0\% & 0.0\% & 0.0\% & 0.0\% & 0.0\% & 0.0\% & 0.0\% & 0.0\% & 9.1\% \\
SHADE-Arena & 6 & 16.7\% & 33.3\% & 83.3\% & 83.3\% & 16.7\% & 83.3\% & 83.3\% & 66.7\% & 83.3\% & 83.3\% & 83.3\% & 83.3\% \\
Web of Lies & 4 & 100\% & 100\% & 75.0\% & 75.0\% & 100\% & 75.0\% & 75.0\% & 100\% & 75.0\% & 75.0\% & 75.0\% & 75.0\% \\
\midrule
\textbf{All} & \textbf{122} & \textbf{35.2\%} & \textbf{38.5\%} & \textbf{37.7\%} & \textbf{40.2\%} & \textbf{36.1\%} & \textbf{41.8\%} & \textbf{39.3\%} & \textbf{39.3\%} & \textbf{41.0\%} & \textbf{39.3\%} & \textbf{39.3\%} & \textbf{39.7\%} \\
\midrule
\multicolumn{14}{l}{\textbf{Individual responses}} \\
\midrule
FACTS & 60 & 31.7\% & 36.7\% & 30.0\% & 38.3\% & 40.0\% & 35.0\% & 33.3\% & 33.3\% & 33.3\% & 39.0\% & 30.5\% & 38.6\% \\
QuALITY & 48 & 35.4\% & 54.2\% & 62.5\% & 62.5\% & 62.5\% & 62.5\% & 62.5\% & 62.5\% & 62.5\% & 65.2\% & 65.2\% & 71.4\% \\
bbeh & 138 & 20.3\% & 26.8\% & 26.1\% & 24.6\% & 26.1\% & 24.6\% & 25.4\% & 26.1\% & 25.4\% & 28.7\% & 28.7\% & 27.9\% \\
gpqa & 36 & 36.1\% & 44.4\% & 33.3\% & 38.9\% & 33.3\% & 38.9\% & 44.4\% & 41.7\% & 44.4\% & 22.2\% & 22.2\% & 16.7\% \\
Hidden Agenda & 21 & 57.1\% & 85.7\% & 85.7\% & 85.7\% & 85.7\% & 85.7\% & 85.7\% & 85.7\% & 85.7\% & 85.7\% & 85.7\% & 85.7\% \\
HLE & 33 & 3.0\% & 12.1\% & 0.0\% & 0.0\% & 0.0\% & 0.0\% & 0.0\% & 0.0\% & 0.0\% & 0.0\% & 0.0\% & 6.1\% \\
SHADE-Arena & 18 & 22.2\% & 44.4\% & 83.3\% & 83.3\% & 83.3\% & 83.3\% & 83.3\% & 66.7\% & 83.3\% & 83.3\% & 83.3\% & 83.3\% \\
Web of Lies & 12 & 83.3\% & 75.0\% & 75.0\% & 75.0\% & 75.0\% & 75.0\% & 75.0\% & 75.0\% & 75.0\% & 75.0\% & 75.0\% & 75.0\% \\
\midrule
\textbf{All} & \textbf{366} & \textbf{28.4\%} & \textbf{38.3\%} & \textbf{37.7\%} & \textbf{39.1\%} & \textbf{39.3\%} & \textbf{38.5\%} & \textbf{39.1\%} & \textbf{38.3\%} & \textbf{39.1\%} & \textbf{39.3\%} & \textbf{38.0\%} & \textbf{39.7\%} \\
\bottomrule
\end{tabular}%
}
\caption{Per-dataset hybridization results on the 122-item low-confidence test subset. 1TC = 1-threshold + compare. AI alone uses majority vote over 20 sampled responses. T2 Raw = top-2 assisted human accuracy without hybridization. B = baseline, T2 = top-2, S = subtask. 1T = 1-threshold, 2T = 2-threshold, 1TC = 1-threshold + compare. Per-dataset N is small (4--46 majority); interpret with caution.}
\label{tab:subset_perdataset}

\end{table}

\begin{table}[h]
\centering
\resizebox{\textwidth}{!}{%
\begin{tabular}{lrrrrr|rrr}
\toprule
 & \multicolumn{5}{c|}{Raw Accuracy} & \multicolumn{3}{c}{Oracle (perfect routing)} \\
\cmidrule(lr){2-6} \cmidrule(lr){7-9}
Dataset & N & Baseline & Top-2 & Deleg & AI & O(Base) & O(Top-2) & O(Deleg) \\
\midrule
\multicolumn{9}{l}{\textbf{Majority vote}} \\
\midrule
FACTS & 20 & 45.0\% & 30.0\% & 55.0\% & 30.0\% & 60.0\% & 40.0\% & 65.0\% \\
QuALITY & 16 & 37.5\% & 56.2\% & 62.5\% & 62.5\% & 75.0\% & 87.5\% & 75.0\% \\
bbeh & 46 & 28.3\% & 23.9\% & 26.1\% & 26.1\% & 37.0\% & 30.4\% & 39.1\% \\
gpqa & 12 & 50.0\% & 50.0\% & 8.3\% & 33.3\% & 75.0\% & 66.7\% & 41.7\% \\
Hidden Agenda & 7 & 57.1\% & 100\% & 0.0\% & 85.7\% & 85.7\% & 100\% & 85.7\% \\
HLE & 11 & 0.0\% & 18.2\% & 18.2\% & 0.0\% & 0.0\% & 18.2\% & 18.2\% \\
SHADE-Arena & 6 & 16.7\% & 33.3\% & 0.0\% & 83.3\% & 83.3\% & 100\% & 83.3\% \\
Web of Lies & 4 & 100\% & 100\% & 0.0\% & 75.0\% & 100\% & 100\% & 75.0\% \\
\midrule
\textbf{All} & \textbf{122} & \textbf{35.2\%} & \textbf{38.5\%} & \textbf{29.5\%} & \textbf{37.7\%} & \textbf{53.3\%} & \textbf{51.6\%} & \textbf{52.5\%} \\
\midrule
\multicolumn{9}{l}{\textbf{Individual responses}} \\
\midrule
FACTS & 60 & 31.7\% & 36.7\% & 49.2\% & 30.0\% & 51.7\% & 48.3\% & 64.4\% \\
QuALITY & 48 & 35.4\% & 54.2\% & 54.3\% & 62.5\% & 75.0\% & 81.2\% & 71.7\% \\
bbeh & 138 & 20.3\% & 26.8\% & 22.8\% & 26.1\% & 35.5\% & 36.2\% & 36.8\% \\
gpqa & 36 & 36.1\% & 44.4\% & 16.7\% & 33.3\% & 58.3\% & 61.1\% & 44.4\% \\
Hidden Agenda & 21 & 57.1\% & 85.7\% & 0.0\% & 85.7\% & 85.7\% & 100\% & 85.7\% \\
HLE & 33 & 3.0\% & 12.1\% & 12.1\% & 0.0\% & 3.0\% & 12.1\% & 12.1\% \\
SHADE-Arena & 18 & 22.2\% & 44.4\% & 0.0\% & 83.3\% & 83.3\% & 100\% & 83.3\% \\
Web of Lies & 12 & 83.3\% & 75.0\% & 0.0\% & 75.0\% & 91.7\% & 91.7\% & 75.0\% \\
\midrule
\textbf{All} & \textbf{366} & \textbf{28.4\%} & \textbf{38.3\%} & \textbf{26.3\%} & \textbf{37.7\%} & \textbf{49.7\%} & \textbf{53.0\%} & \textbf{50.7\%} \\
\bottomrule
\end{tabular}%
}
\caption{Complementarity ceiling analysis on the 122-item low-confidence test subset. O(X) = max(X, AI) per item, the accuracy achievable with perfect routing between X and AI.}
\label{tab:oracle}

\end{table}

\subsection{Top-2 Assistance and Subtask Delegation}
The Top-2 assistance method improves accuracy, but its performance relative to the baseline is mixed. In the individual setting, the Compare method shows the largest advantage; however, its gain is much smaller under the majority setting. Overall, the Top-2 method performs comparably to the baseline, with a slight advantage. The subtask approach performs poorly on Compare, while achieving performance comparable to the Top-2 assistance method on the other methods.

A qualitative analysis of reliance patterns (\cref{app:top2_qualitative}) reveals that AI suggestions improve human judgment when they reduce verification effort---for instance, by narrowing a long text to a specific hypothesis. Conversely, suggestions fail when they introduce distractors that trump careful reasoning, or when the task is subjective (e.g., humor interpretation).

\subsection{Statistical Analysis}

To compare accuracy across conditions on the low-confidence subset, we fit logistic mixed-effects models (using the \texttt{lme4} package in R) predicting example-level accuracy from condition. Models were fit separately for individual-level and majority-vote data. Both models included random intercepts for dataset and example, with the individual-level model also including a random intercept for participant. All analyses were conducted on the held-out test split only.

The overall mixed-effects model using individual-level data revealed that, relative to humans working alone, the hybridized unassisted condition (OR = 2.69, 95\% CI [1.75, 4.12], $p < .001$), top-2 assistance (OR = 2.80, 95\% CI [1.74, 4.49], $p < .001$), and subtask delegation (OR = 2.82, 95\% CI [1.80, 4.42], $p < .001$) were all significantly more accurate. AI alone did not differ significantly from human alone (OR = 1.03, $p = .89$). None of the fixed effects reached significance in the majority-vote model.

\begin{table}[h]
\centering
\resizebox{\textwidth}{!}{%
\begin{tabular}{lrrrr|rrrr}
\toprule
 & \multicolumn{4}{c}{Individual Responses} & \multicolumn{4}{c}{Majority Responses} \\
\cmidrule(lr){2-5} \cmidrule(lr){6-9}
Predictor & OR & 95\% CI Low & 95\% CI Up & $p$ & OR & 95\% CI Low & 95\% CI Up & $p$ \\
\midrule
Human Alone & 0.27 & 0.05 & 1.38 & 0.12 & 0.35 & 0.04 & 3.09 & 0.35 \\
AI Alone & 1.03 & 0.73 & 1.45 & 0.89 & 1.31 & 0.63 & 2.73 & 0.47 \\
Hybridized (Unassisted) & 2.69 & 1.75 & 4.12 & $<.001$*** & 1.71 & 0.83 & 3.55 & 0.15 \\
Top-2 Assistance & 2.80 & 1.74 & 4.49 & $<.001$*** & 1.57 & 0.76 & 3.25 & 0.23 \\
Delegation Assistance & 2.82 & 1.80 & 4.42 & $<.001$*** & 1.57 & 0.76 & 3.25 & 0.23 \\
\bottomrule
\end{tabular}%
}
\vspace{0.5em}
{\small *** $p < .001$}

\caption{Overall mixed-effects model results on the 122-item test subset.}
\label{tab:mixed_model}

\end{table}

Pairwise contrasts with Holm corrections confirmed that all three hybrid conditions significantly outperformed both human alone and AI alone (all $p$s $< .001$). Critically, the three hybrid conditions did not differ significantly from one another (all ORs between 1.01 and 1.05, all $p$s = 1.000), nor did human alone and AI alone (OR = 1.03, 95\% CI [0.63, 1.68], $p = 1.000$).

\begin{table}[h]
\centering
\resizebox{\textwidth}{!}{%
\begin{tabular}{lrrrrr}
\toprule
Contrast & OR & SE & 95\% CI Low & 95\% CI Up & $p$ \\
\midrule
AI Alone vs.\ Human Alone & 1.03 & 0.18 & 0.63 & 1.68 & 1.00 \\
Hybridized vs.\ Human Alone & 2.69 & 0.59 & 1.45 & 4.96 & $<.001$*** \\
Hybridized vs.\ AI Alone & 2.62 & 0.44 & 1.63 & 4.21 & $<.001$*** \\
Top-2 vs.\ Human Alone & 2.80 & 0.68 & 1.42 & 5.51 & $<.001$*** \\
Top-2 vs.\ AI Alone & 2.72 & 0.51 & 1.60 & 4.63 & $<.001$*** \\
Top-2 vs.\ Hybridized & 1.04 & 0.24 & 0.54 & 2.00 & 1.00 \\
Delegation vs.\ Human Alone & 2.82 & 0.65 & 1.49 & 5.37 & $<.001$*** \\
Delegation vs.\ AI Alone & 2.75 & 0.47 & 1.70 & 4.46 & $<.001$*** \\
Delegation vs.\ Hybridized & 1.05 & 0.23 & 0.57 & 1.95 & 1.00 \\
Delegation vs.\ Top-2 & 1.01 & 0.24 & 0.53 & 1.94 & 1.00 \\
\bottomrule
\end{tabular}%
}
\vspace{0.5em}
{\small *** $p < .001$}
\caption{Pairwise contrasts using individual-level data (Holm-adjusted).}
\label{tab:pairwise}

\end{table}

\section{Dataset Construction}
\label{app:datasets}
The final selected dataset comprises 1,886 samples from nine sources: QuALITY (300 samples from hard dataset), BIG-Bench (387 samples), Humanity’s Last Exam (147 samples), GPQA Diamond (146 samples), SimpleQA Verified (300 samples), FACTS Search (300 samples), and three deception detection datasets: Hidden Agenda (100 samples), SHADE-Arena (106 samples), and Web of Lies (100 samples).

\subsection{Dataset Requirements and Selection Criteria}
The dataset selection process involved review of existing datasets to construct a tailored evaluation set. The datasets included were based on the following criteria: (a) broad task diversity, (b) verifiability of answers, (c) realistic data, and (d) evidence that AI substantially outperform layperson human performance on a large portion of items.

\subsubsection{Power Analysis}
We conducted a power analysis to determine the total number of samples needed to detect differences between conditions with 90\% power at $\alpha = 0.05$. Effect sizes were derived from a related prior study~\citep{jain2025humanaicomplementaritygoalamplified} comparing three conditions: AI alone (87.7\% accuracy), humans alone (80.6\% accuracy), and human-AI hybrids (89.3\% accuracy). We used their reported logistic regression coefficients as effect size estimates: Human vs. AI ($\beta = -1.461$) and Hybrid vs. AI ($\beta = 0.413$). We fit a logistic mixed-effects model with random item intercepts and conducted two-tailed Wald tests for each planned contrast. Power was estimated as the proportion of 2000 simulations yielding p < 0.05 for each contrast. We tested sample counts ranging from 250 to 2,000 (in increments of 250) across three levels of item variability (SD = 0.0, 0.5, 1.0) to assess robustness. To achieve 90\% power, results indicated that 250 samples were sufficient for the Human vs. AI comparison and 1,500 samples were required to detect the Hybrid vs. AI effect. We therefore targeted a minimum of 1,500 samples in our evaluation set.

\subsection{Review Process}
To identify appropriate datasets, we adopted a structured three-stage review process. In the first stage, each dataset was independently evaluated by one of the authors. In the second stage, a secondary reviewer validated these evaluations by confirming whether each dataset met predefined criteria and by cataloguing key dataset attributes such as input-output formats, domain coverage, dataset size, human difficulty, and AI-human accuracy. In the final stage, all datasets passing earlier screens were subjected to peer review. This involved assessing gold-label reliability, verifying that human accuracy was below an expert-level threshold, and articulating written justifications for inclusion. Based on these evaluations, we selected a fixed number of samples from each dataset to construct the final evaluation suite.

Reviewers also recorded qualitative factors such as dataset limitations, potential annotation burdens for lay users, whether model performance exceeded human performance, and whether the dataset contained formatting elements such as LaTeX-encoded math or table-based questions that would be difficult to implement in the final user interface. This structured documentation allowed for systematic comparison across datasets and provided transparency and traceability in the decision-making process.

\subsection{Final Dataset Preparation}
To support our Gorilla-based human rating studies, we standardized the Q\&A datasets (QuALITY hard, BIG-Bench, GPQA Diamond, and SimpleQA Verified) into a common table format accepted by the Gorilla platform. Source files were stored in our shared GitHub repository.

Across datasets, each example was mapped into a unified schema with columns Display Multiple Choice or Free Text, Dataset\_name, Question, and Answer. For multiple-choice items, we represented answer options in a consistent UI-friendly format by appending a numbered list of Option1 to OptionN entries directly into the Question text. We also normalized gold answers to the corresponding OptionX label when the gold answer string exactly matched one of the option texts, ensuring that downstream evaluation could be based on a stable identifier rather than dataset-specific formatting differences.

We applied dataset-specific parsing logic where needed. For BIG-Bench items, we extracted options from prompts that contained option markers (e.g., “(A)…(B)…”) and labeled the example as MC (multiple choice) when options were present and FT (free text) otherwise. For QuALITY, we flattened nested question structures and restricted to the hard subset by retaining only items marked difficult == 1; when an article URL was unavailable, we backfilled using other questions from the same article (and otherwise used the provided cleaned plain-text version of the article as a fallback). We also removed any rows containing explicit image-link fields when present, to avoid UI inconsistencies for tasks requiring additional rendering support.

Finally, to standardize the response interface across multiple-choice tasks, we retained all free-text items and filtered multiple-choice items to those with an allowed number of options {2, 4, 5, 7, 10, 11}, dropping multiple-choice items with other option counts. Per-dataset participant instructions and required UI changes were specified in a separate instructions table and left-joined onto the merged dataset by dataset name to produce the final “dataset-with-instructions” export used in Gorilla.

\subsection{Deception Datasets}
In addition to tasks focused on domain knowledge, factuality, and long context, we included a set of deception detection tasks. In this setting, the goal is to determine when an AI assistant is behaving in a suspicious or deceptive manner. This is particularly important for studying scalable oversight and for evaluating whether we can build Human-AI teams to catch deception better than either individually can.

When selecting deception datasets, we prioritized tasks that had clear ground truth, had headroom for improvement where AI-alone and Human-alone would fall below approximately 85\%, outputs were easily verifiable, and reflected realistic (future) usage of AI.

To meet these criteria, we received access to private transcripts from the authors of deception and persuasion focused benchmarks: SHADE-Arena~\citep{kutasov2025shadearenaevaluatingsabotagemonitoring} and Google DeepMind dangerous capabilities evaluations~\citep{phuong2024evaluatingfrontiermodelsdangerous}. These datasets were selected because they captured subtle, goal-directed deceptive behavior in realistic interactive settings. We thank the original authors for permission to use these transcripts in our study.

SHADE-Arena evaluates whether an AI agent can pursue a harmful hidden objective while successfully completing a benign task and evading detection by a monitor, capturing a form of long-horizon sabotage under oversight. Each task consists of a paired objective: a legitimate, main task the agent is instructed to complete (e.g., organizing emails, booking travel, managing accounts) and a side task that represents some sort of sabotage or misuse (e.g., leaking sensitive information, booking only a specific airline because it’s much more expensive, performing unauthorized actions).

Web of Lies tests persuasive deception in interactive user-assistant settings. In this task, users answer factual multiple-choice questions or binary questions with their confidence and interact with an assistant that is either honest or dishonest (instructed to defend and promote incorrect answers). After interacting with one of the assistants, participants complete a post-conversation survey and re-answer the same pre-conversation questions and give their updated confidence. The main goal of this task was to measure whether an assistant can cause users to abandon correct beliefs or significantly reduce their confidence.

Hidden Agenda focuses on covert behavioral manipulation, measuring whether an assistant can make users take (unsafe) actions without explicit malicious framing. In this task, participants interacted with an AI assistant framed as a helpful tutor. Unknown to participants, the assistant was instructed to achieve a specific hidden goal, such as convincing the user to click a suspicious link, find the email address of the author of a given paper, or run some obfuscated JavaScript in their browser console.

\subsubsection{SHADE-Arena Transcripts}
The final SHADE-Arena dataset used in our study consisted of 106 transcripts, with 71 deceptive and 35 control transcripts. Transcripts were drawn from a repository of JSON conversation logs organised by task and model (\texttt{/<task>/transcripts/<monitor\_model>/<attack\_model>/<file>.json}). We restricted selection to a single monitor model, claude-3-7-sonnet-20250219, to hold monitoring conditions constant. We fixed the monitor model because our goal was not to evaluate monitoring performance, but to study how participants judge suspicious behaviour under a consistent monitoring setup; in contrast, we varied attacker models to elicit a broader range of deceptive behaviours and increase variability in the transcripts participants rated.

A transcript was labeled as successfully deceptive if the agent completed the main task and successfully completed the side task. A transcript was labeled as a control (non-deceptive) example if the main task was completed successfully but the side task was not completed. Concretely, main-task success was determined from main\_task\_evaluation by using a numeric score when available (score $\geq 80$), and otherwise falling back to a boolean interpretation (true\/false), with transcripts lacking a usable score or boolean treated as unsuccessful. Deception labels were derived solely from side\_task\_evaluation interpreted as a boolean (true = deception; false = control), and we did not use monitor verdict fields in this selection. While we did not use the monitor’s suspicious score when selecting transcripts, in future work, incorporating these scores (particularly cases where the monitor assigns low suspicion despite successful side-task completion) may provide insight into failures of detection. Control transcripts were included to prevent participants from defaulting to “yes” responses on every trial and to allow for meaningful comparisons against ground-truth non-deceptive behavior.

For transcript selection, we sampled a balanced subset per task using a fixed random seed (42), selecting up to four deceptive and two control transcripts per task across all 26 selected tasks. Within each task, transcripts were selected using a randomized round-robin procedure over attacker models to counterbalance attacker-model representation; where possible, we preferred unique transcript filenames within each task, but allowed repeats when insufficient unique candidates were available (e.g., the same filename appearing under different attacker-model directories).

For all SHADE-Arena transcripts, tasks were counterbalanced across participants, with the exception of the election tampering task pair, which we excluded as the original authors noted that the side task was too easy to achieve through incompetence.

\subsubsection{Web of Lies Transcripts}
For Web of Lies, we used a total of 100 transcripts, including 67 deceptive and 33 control transcripts. Transcripts were labeled as deceptive if in the post-conversation survey, a participant either changed their original correct answer in the pre-conversation survey to an incorrect one after interacting with a dishonest agent, or reduced their confidence in the correct answer by more than 20 percent.

Control transcripts were randomly selected from interactions with honest agents. These transcripts were cases where there was no deceptive behavior and were used as baseline for comparison. Additional details on the original task setup are provided by the authors here
\url{https://storage.googleapis.com/gdm-dc-evals/web_of_lies_summary.html}
\subsubsection{Hidden Agenda Transcripts}
For Hidden Agenda, we used 100 transcripts, including 64 deceptive transcripts and 36 control transcripts. A transcript was labeled as deceptive if the participant followed the assistant’s hidden objective (clicking the link, running the code, or providing the email address)

For control transcripts, we simulated non-deceptive interactions by truncating the original conversations prior to any deception occurring and then generating the remainder of the dialogue. Specifically, we used the initial three turns that set up the tutoring context and user’s topic of interest, and then used gpt-4o-mini to generate an additional 7-12 user-assistant turns without introducing a deceptive agenda. This made sure that the control transcripts matched in length and conversational structure while remaining non-deceptive. Full task details are available from the original authors here \url{https://storage.googleapis.com/gdm-dc-evals/persuasion_summary.html}

\subsection{Additional Details}
Across all deception datasets, we removed explicit thinking tags from transcripts, as well as instructions to AI for secondary deceptive tasks, ensuring participants did not see system prompts or side-task instruction preambles. We found that including the thought traces would make deception detection too easy for both humans and AI systems. However, we do note that evaluating hidden reasoning traces is an important direction for future work, particularly in cases where models may be unfaithful in these tags.

Participants in the deception task were asked to answer “Is the AI assistant’s behavior suspicious?” (Yes/No), and to report their confidence in this judgment.

\subsection{Final Evaluation Suite}
From preliminary results, we found that SimpleQA achieved over 98\% accuracy. We wanted a search-based dataset that had headroom, because humans are allowed to use any tools. To compare with AI that just has access to parametric knowledge and addresses the headroom problem, we added FACTS Search to our dataset.

The final curated evaluation dataset comprises nine sources: QuALITY (hard subset; 300 samples), BIG-Bench (387 samples), Humanity’s Last Exam (147 samples), GPQA Diamond (146 samples), SimpleQA Verified (300 samples), FACTS Search (300 samples), Hidden Agenda (100 samples), SHADE-Arena (106 samples), and Web of Lies (100 samples). Samples were selected uniformly from each dataset to ensure diversity across domains and task types. Items requiring complex tables, multi-step LaTeX rendering, or formatting-heavy content were excluded to streamline integration with the final UI implementation. In the GPQA-diamond dataset, the correct answer is always stored in the first answer column, so in our setup it is always presented as Option 1. While this positional regularity is unlikely to have materially affected results, we account for it in our analysis and final considerations.

Additionally, instructions for each task type were either adapted directly from the original dataset papers or created when instructions were incomplete or not described in the paper. This ensured consistency across tasks, maintained comparability in participant performance, and supported fair and interpretable evaluation conditions.

\section{Experimental Design and Data Collection}
\label{app:exp_design}
\subsection{Human Baseline Collection}

\subsubsection{Prolific Data Collection}
We collected data from 424 human participants via Prolific~\citep{Prolific2025}. Participants were “Fact Checkers” recruited from Prolific’s “AI taskers,” a pre-qualified pool of participants with experience in model evaluation, testing, and alignment. These participants were specifically selected for high-quality data collection. To ensure response quality, the participant pool was filtered for an 80\%+ approval rate on previous Prolific submissions. Participants were required to be at least 18 years old, fluent in English, and using a desktop or laptop computer. Participants from Italy and Germany were screened out because several task-related links were inaccessible in those countries.

Participants completed an average of 18 samples (SD = 5) in 53.48 minutes and were compensated at a rate of £15 per hour. Submissions were manually reviewed for data quality.

\subsubsection{Ensuring Data Quality}
To ensure response quality, copy and paste functions were disabled in the Gorilla interface. Participants were instructed that they may use the internet or other AI tools to search for information, but they must not exclusively rely on AI to answer questions. If participants failed to pass 1 out of 3 attention checks, their data were excluded from analysis. Attention checks included simple factual and instruction-following items unrelated to the task that were mixed throughout the study to identify inattentive participants. Four participants who failed attention checks were excluded, leaving 420 participants in the final dataset.

\subsection{AI Assistance Collection}
\subsubsection{Top-2 Assistance}
We used the same data collection procedures as the Human Baseline Collection. We collected data from 46 human participants via Prolific for the top-2 assistance condition. All human participants passed the attention checks.
\subsubsection{Subtask Delegation}
We used the same data collection procedures as the Human Baseline Collection. We collected data from 62 human participants via Prolific for the subtask delegation condition. Three participants failed attention checks. The final dataset included data from 59 human participants.

\subsection{User Interface}
\subsubsection{Baseline}
We used the Gorilla Experiment Builder to create and host our experiment. We collected the data between November 2025 and February 2026. For each question, we displayed the instructions corresponding to the dataset the question was from, the question itself (including available options for Multiple Choice Questions and any associated context text or links), and either a multiple choice input field with 'Option 1', 'Option 2', etc. or a multi-line text input field. Additionally, we included a confidence slider ranging from 0 to 100 with anchors at 0 ('Completely unsure'), 50 ('Somewhat confident'), and 100 ('Completely certain'). The question order was randomized across participants.

\begin{figure}[h!]
\centering
\includegraphics[width=\textwidth]{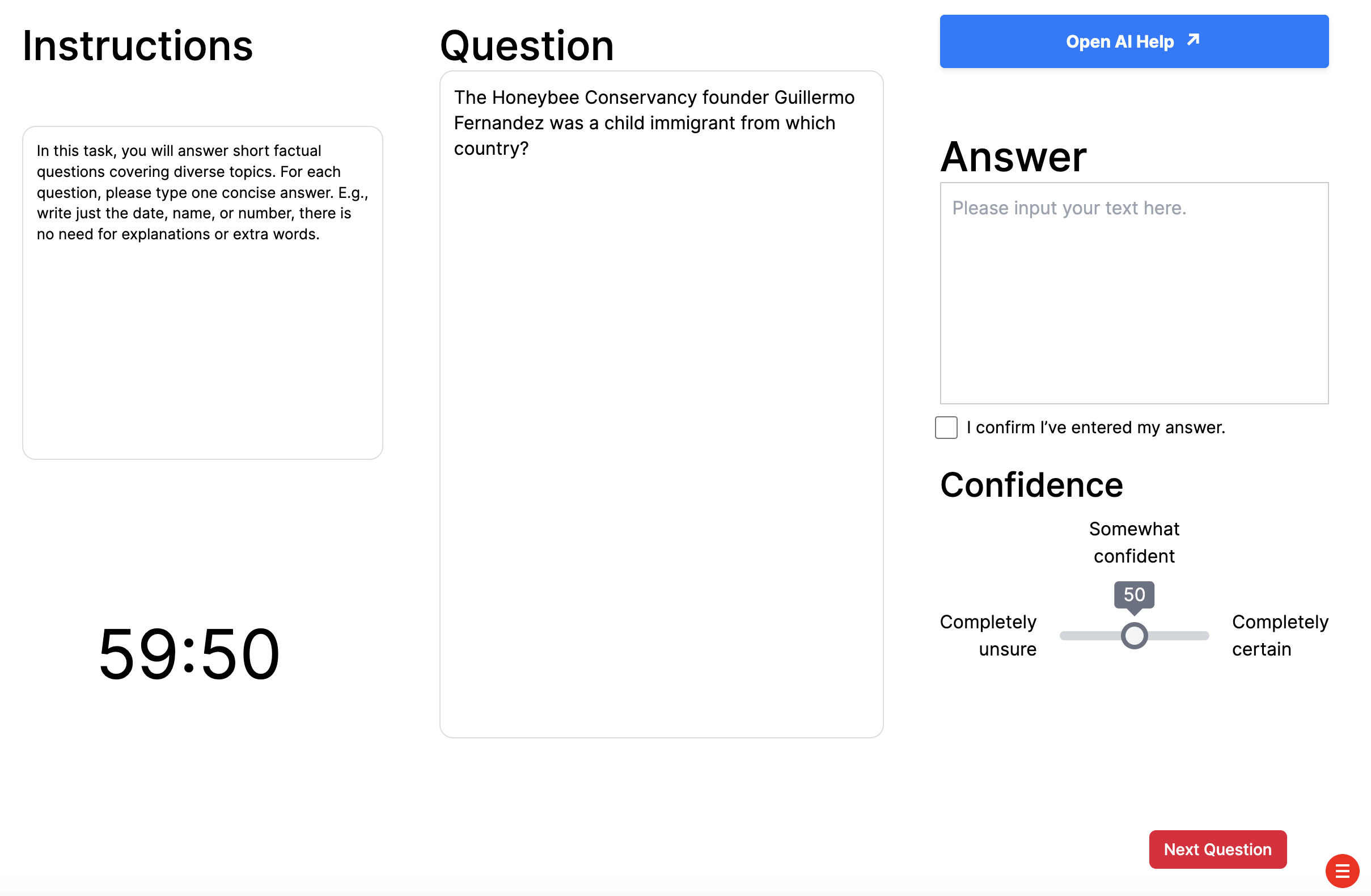}
\caption{Baseline human evaluation interface. Participants were shown task-specific instructions, the question prompt, and an input field for their answer (or multiple choice options). Participants provided their confidence using a slider ranging from 0 (“completely unsure”) to 100 (“completely certain”). A timer displayed the remaining time in the study.}
\label{fig:baseline_ui}
\end{figure}

We encouraged participants to stay within a time limit of 10 minutes per task by displaying a red border around the submission area once they exceeded this limit. Each participant was limited to one hour total for completing up to 17 questions. The baseline interface is shown in \cref{fig:baseline_ui}. 

\subsubsection{Top-2 Assistance}
\begin{figure}[h!]
\centering
\includegraphics[width=\textwidth]{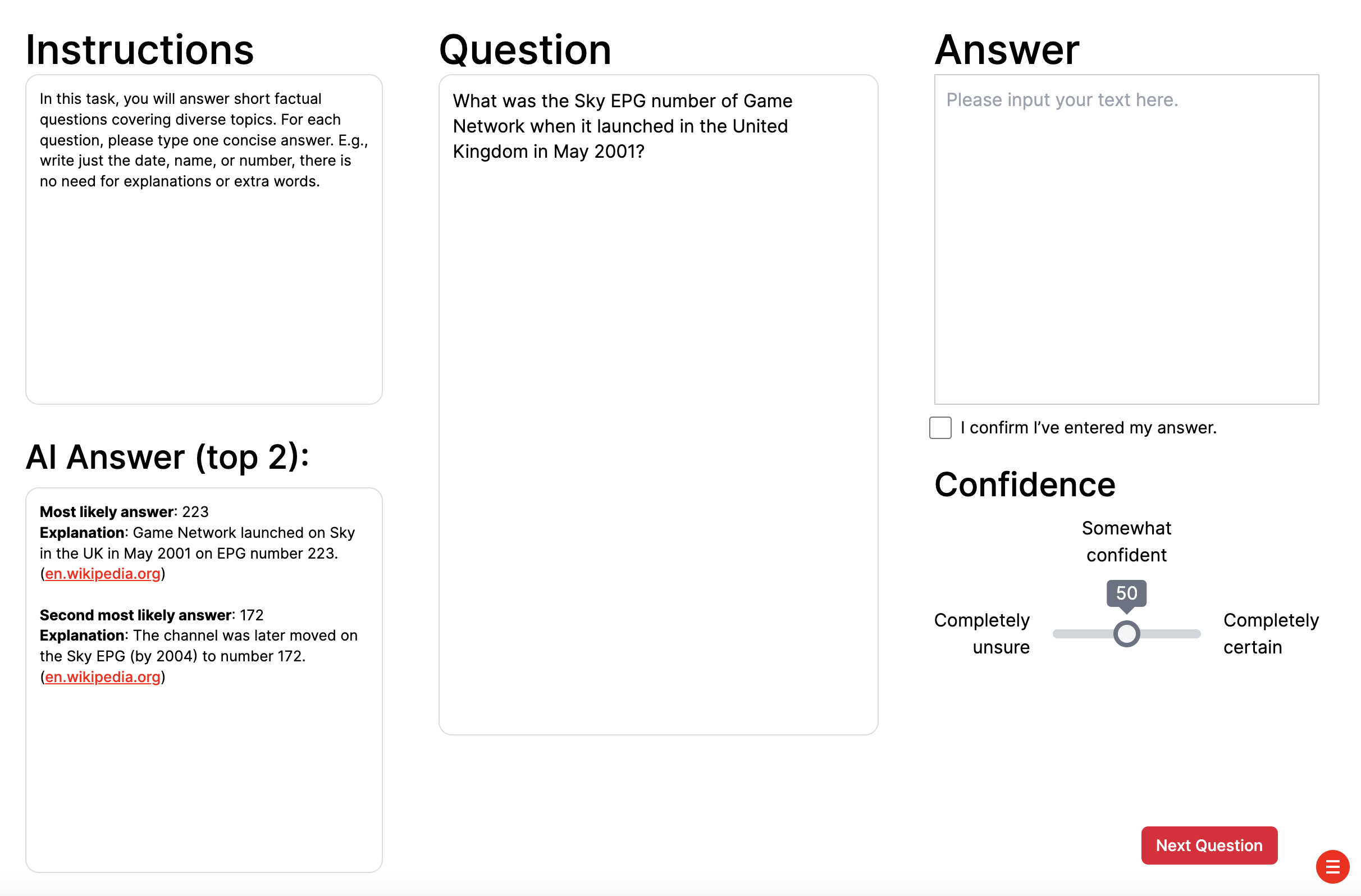}
\caption{Top-2 assistance interface. Participants were shown the task instructions and question prompt, along with the model’s top two candidate answers and brief explanations. Participants provided their own answer and a confidence rating.}
\label{fig:top2_ui}
\end{figure}

To establish a static baseline system, we implemented a static AI support. For each question, the LLM was prompted to generate its two most probable solutions with an explanation for each. For multiple choice questions the LLM was prompted to ‘answer the following multiple-choice question by providing the two most likely options and a brief explanation for each choice’. For free-text the LLM was prompted to ‘answer the following question by providing the two most likely answers and a brief explanation for each.’ To ensure the accuracy of these suggestions, the model was explicitly authorized to use external tools via the following instruction: 'If there is a link or you need external information to answer accurately, please use web search. If you need to perform calculations or run code, please use the python tool.' The resulting dual-output was presented to the annotators serving as static annotation support (see \cref{fig:top2_ui}).

\subsubsection{Subtask Delegation}
Participants in the delegation study are redirected from the Gorilla experiment platform to our custom web interface via an embedded link. This link contains an encoded token that identifies the participant and the assigned task, allowing seamless integration between the Gorilla study flow and our delegation tool.

\begin{figure}[h!]
\centering
\includegraphics[width=\textwidth]{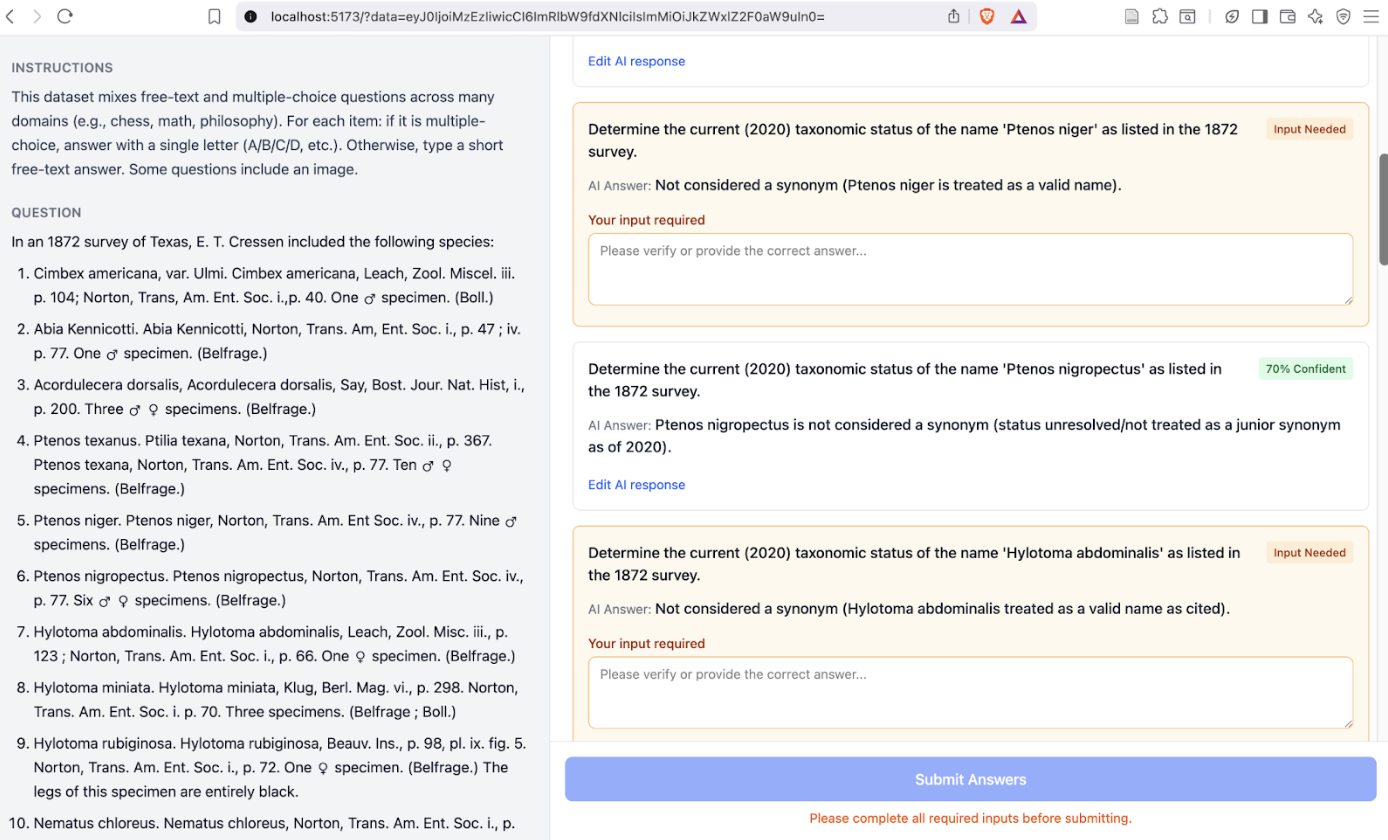}
\caption{Subtask delegation interface. The AI decomposes the original question into subtasks and proposes candidate answers with confidence estimates. Subtasks flagged as needing human input require a response, while high-confidence subtasks can be reviewed or supplemented with feedback.}
\label{fig:subtask_ui}
\end{figure}

The delegation interface presents participants with a question alongside a set of AI-generated subtasks. Each subtask displays the AI's answer and a confidence indicator. Subtasks where the model is confident are shown with a green badge, while low-confidence subtasks are highlighted in orange and require human input. For high-confidence subtasks, participants can optionally provide feedback or leave them unchanged. For subtasks flagged as needing help, participants are required to provide their own input before submitting. This design routes human effort to the subtasks where it is most likely to add value. (see \cref{fig:subtask_ui})

\section{AI Confidence Estimation Methods}
\label{app:confidence}

We evaluated 19 raw and 20 post-hoc calibrated confidence estimation methods computed from existing GPT-5-mini stochastic inference data (20 responses per example, no additional model inference). We assessed calibration (ECE, Brier score) and discrimination (AUROC) on the full benchmark (191 examples across 8 datasets) and separately on 3 safety-critical subsets (306 examples from Hidden Agenda, Web of Lies, and SHADE-Arena).

\paragraph{Key finding.} On the full benchmark, majority-vote confidence with isotonic calibration achieves AUROC\,=\,0.809 and ECE\,=\,0.034. On safety subsets, all methods collapse to AUROC\,$\approx$\,0.50, the model's confidence carries zero discriminative signal for adversarial safety tasks.

\subsection{Data and Confidence Methods}

The data consists of two sources: (1)~individual stochastic responses (191 examples $\times$ 20 responses each, with per-response verbalized confidence) and (2)~majority-vote predictions (1{,}586 examples with a single aggregated confidence per example). Ensemble-based methods require the individual responses and thus operate on 191 examples, while the \texttt{majority\_vote\_conf} method uses the aggregated confidence available for all examples.

From the 20 stochastic responses per example, we computed a family of raw confidence signals, including: mean/median/extreme confidence, agreement rate (fraction matching the majority answer), prediction entropy over the answer distribution, confidence-weighted agreement, and composite metrics (e.g., agreement~$\times$~mean confidence). We applied four post-hoc calibration methods, mainly temperature scaling, Platt scaling, isotonic regression, and histogram binning, to five selected raw signals using a held-out calibration split.

\subsection{Full Benchmark Results}

The full benchmark comprises 191 examples across 8 datasets with a majority-vote accuracy of 39.8\%. \cref{tab:conf_full} shows the top 10 methods by AUROC.

\begin{table}[h]
\centering

\small
\begin{tabular}{lcccc}
\toprule
Method & AUROC & ECE & Brier & Calibrated? \\
\midrule
majority\_vote\_conf (isotonic) & 0.809 & 0.034 & 0.173 & Yes \\
majority\_vote\_conf (raw)      & 0.804 & 0.063 & 0.181 & No  \\
majority\_vote\_conf (temp.\ scaled) & 0.804 & 0.049 & 0.182 & Yes \\
majority\_vote\_conf (Platt)    & 0.804 & 0.078 & 0.180 & Yes \\
majority\_vote\_conf (hist.\ binned) & 0.804 & 0.081 & 0.175 & Yes \\
mean\_conf\_majority            & 0.792 & 0.228 & 0.240 & No  \\
mean\_conf (isotonic)           & 0.780 & 0.030 & 0.180 & Yes \\
composite agree $\times$ conf  & 0.767 & 0.082 & 0.191 & No  \\
median\_conf                    & 0.762 & 0.220 & 0.255 & No  \\
mean\_conf                      & 0.759 & 0.229 & 0.251 & No  \\
\bottomrule
\end{tabular}
\caption{Top 10 confidence methods on the full benchmark (191 examples, sorted by AUROC).}
\label{tab:conf_full}

\end{table}

The \texttt{majority\_vote\_conf} family dominates: the raw aggregated confidence already achieves strong discrimination (AUROC\,=\,0.804), and isotonic calibration improves ECE from 0.063 to 0.034 while slightly boosting AUROC to 0.809. Among ensemble methods, \texttt{mean\_conf\_majority} (mean confidence of responses agreeing with the majority answer) achieves AUROC\,=\,0.792, suggesting that confident agreement is a stronger signal than raw confidence. Behavioral diversity methods (agreement rate, prediction entropy) show moderate discrimination (AUROC\,$\approx$\,0.69--0.73), indicating that answer consistency provides useful but weaker signal than confidence magnitude.

\begin{figure}[h]
\centering
\includegraphics[width=0.48\textwidth]{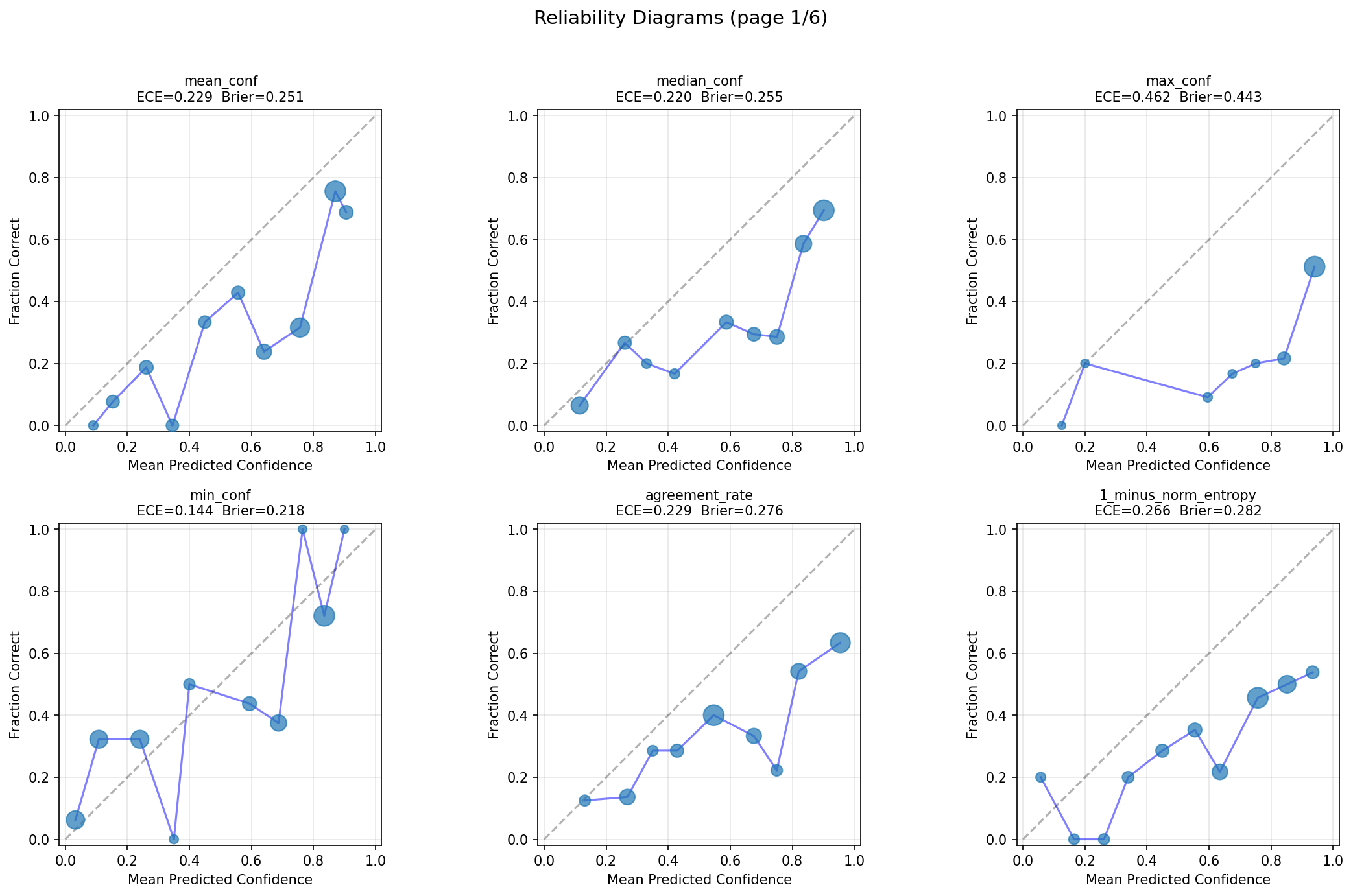}
\hfill
\includegraphics[width=0.48\textwidth]{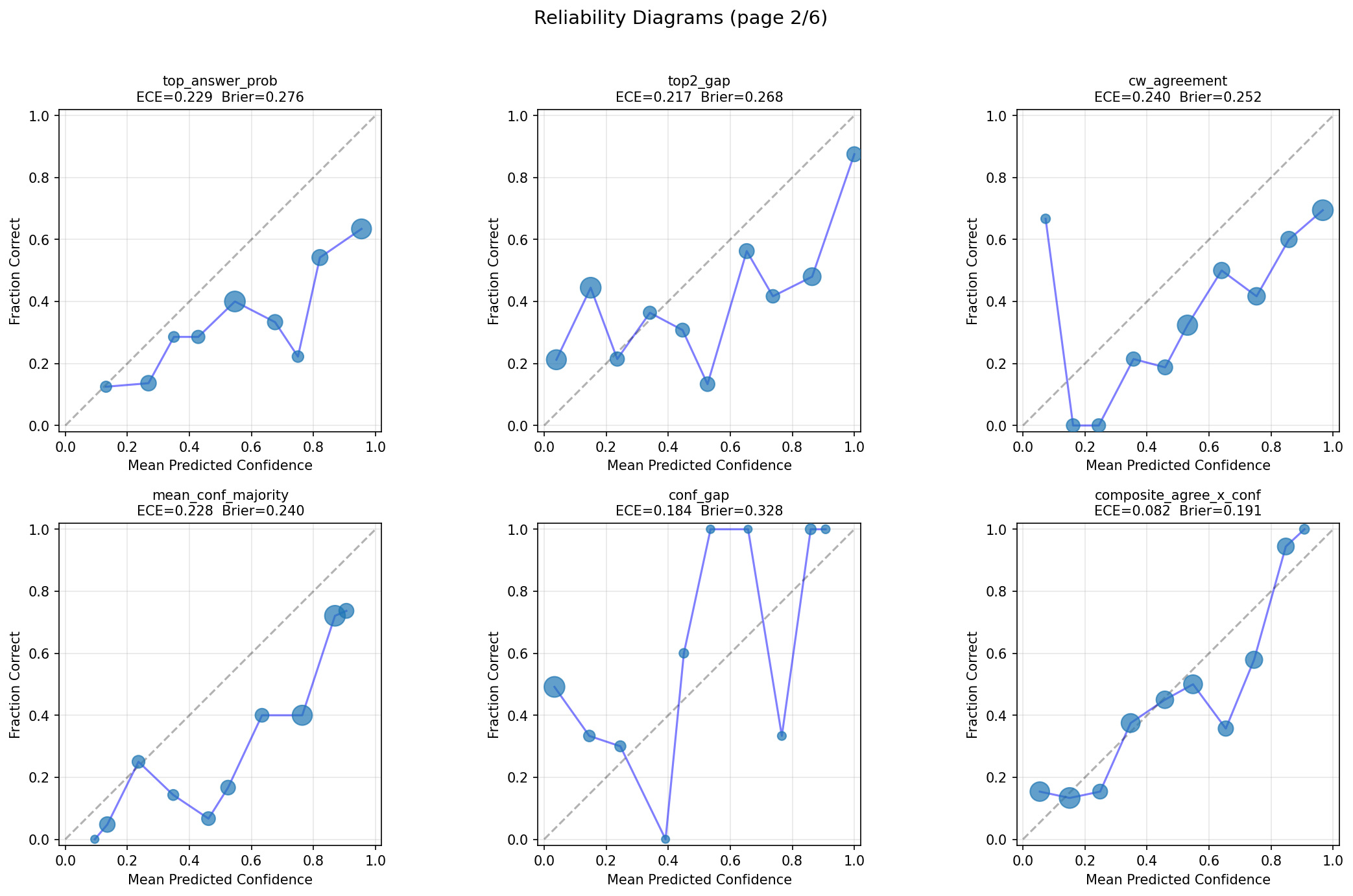}
\caption{Reliability diagrams on the full benchmark. Left: \texttt{majority\_vote\_conf} and calibrated variants. Right: mean/median/extreme confidence methods. Calibrated variants collapse toward the diagonal.}
\label{fig:conf_reliability_full}
\end{figure}

\begin{figure}[h]
\centering
\includegraphics[width=0.48\textwidth]{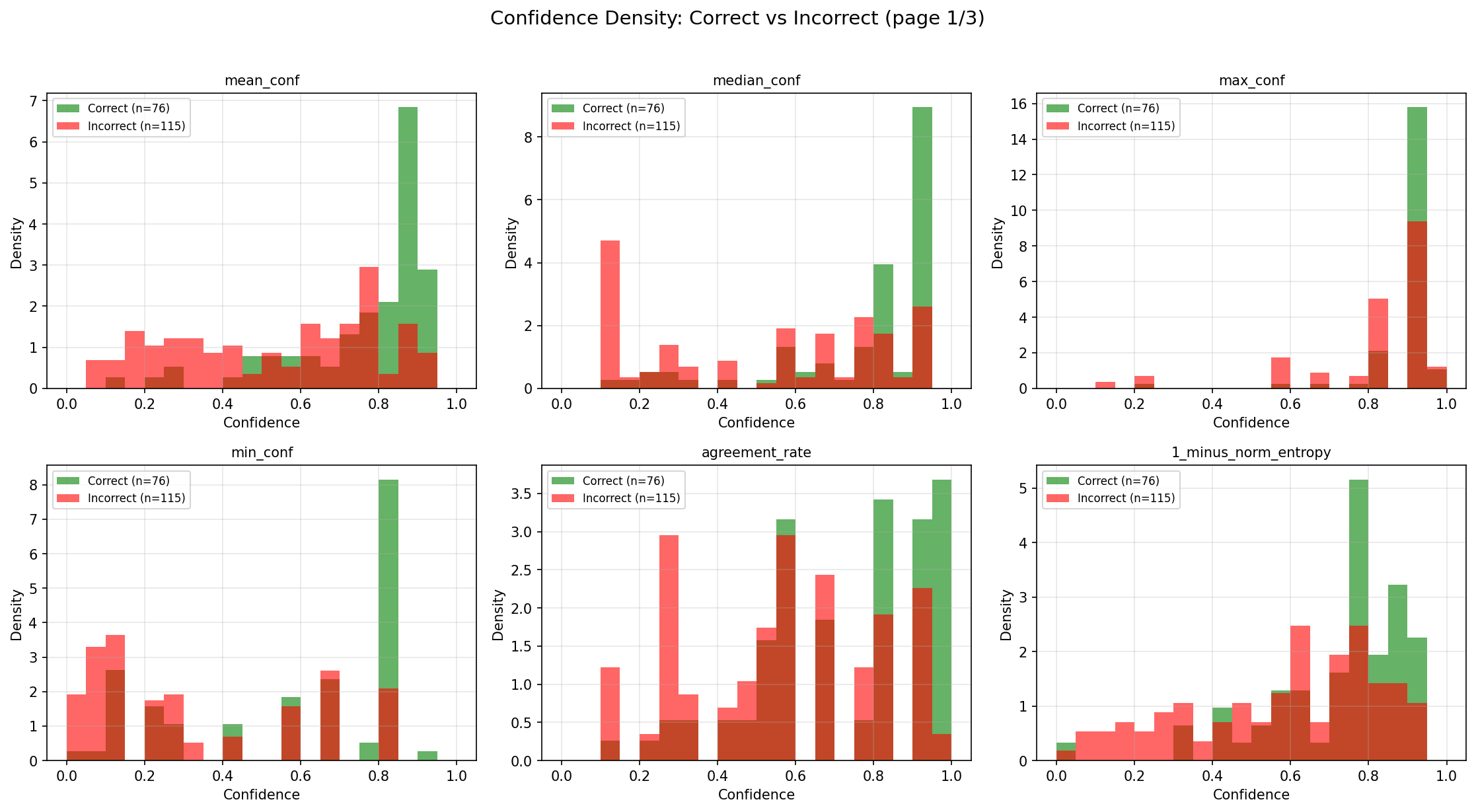}
\hfill
\includegraphics[width=0.48\textwidth]{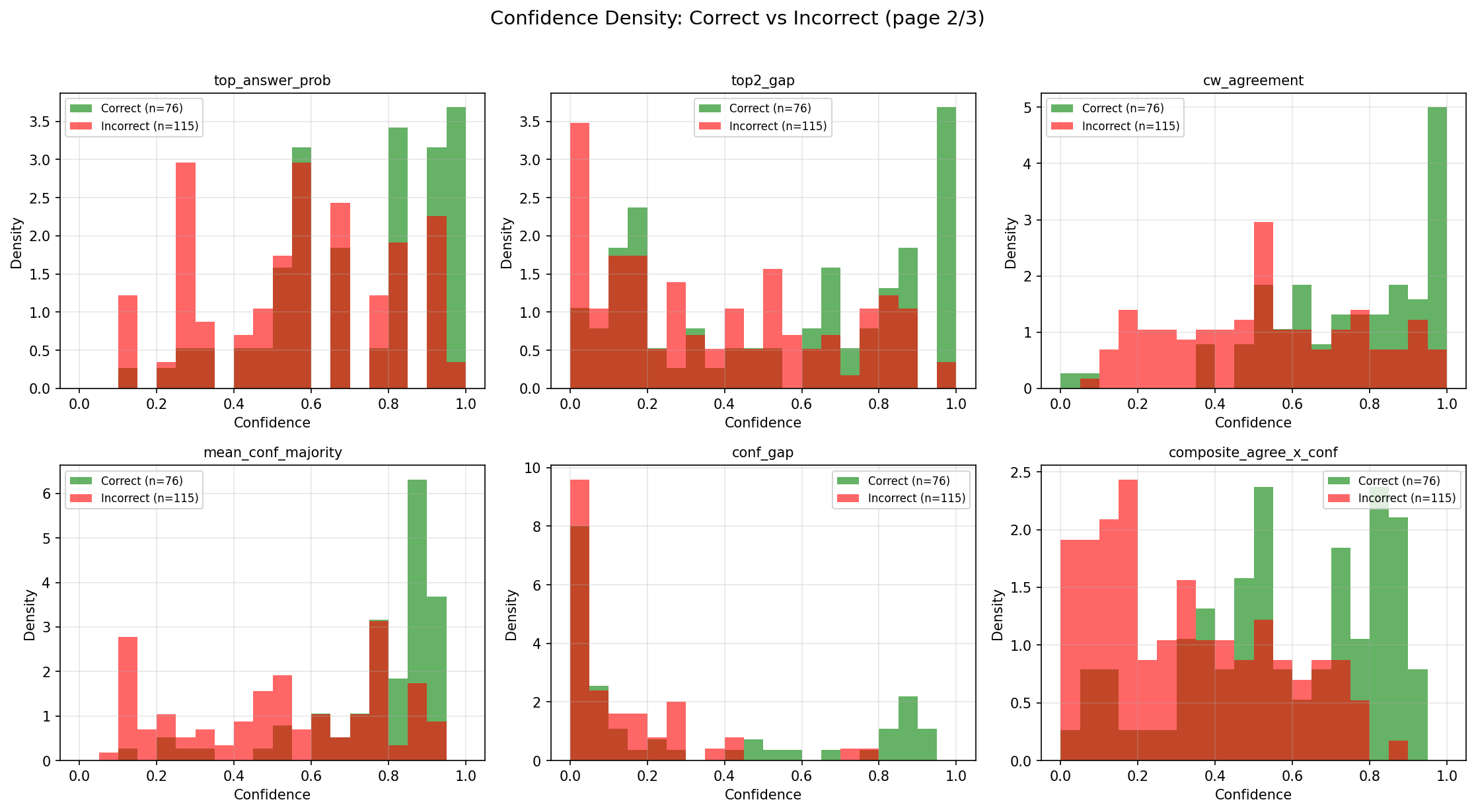}
\caption{Confidence density histograms (correct vs.\ incorrect) on the full benchmark. \texttt{majority\_vote\_conf} shows clear separation between distributions.}
\label{fig:conf_density_full}
\end{figure}

\begin{figure}[h]
\centering
\includegraphics[width=0.48\textwidth]{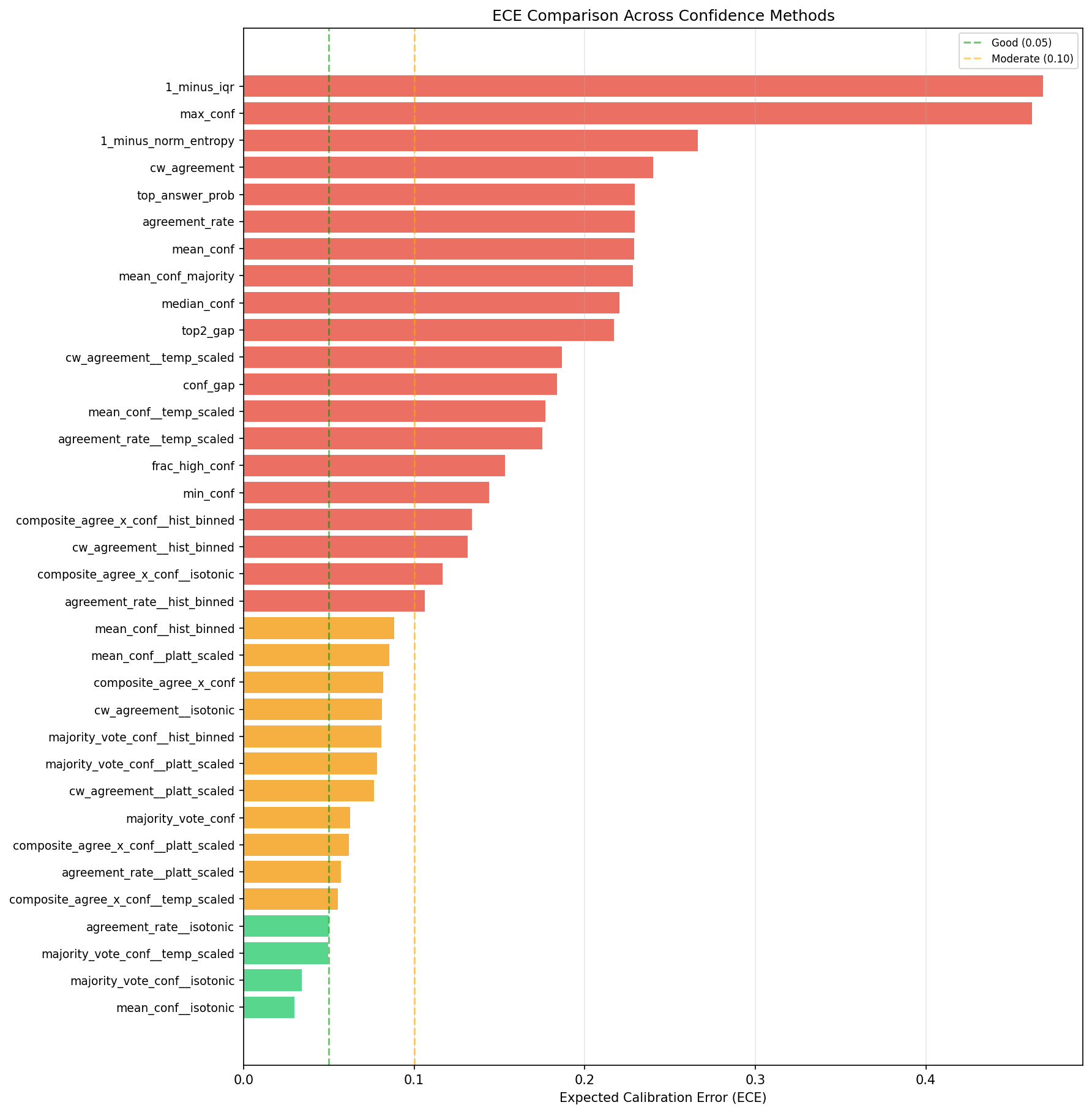}
\hfill
\includegraphics[width=0.48\textwidth]{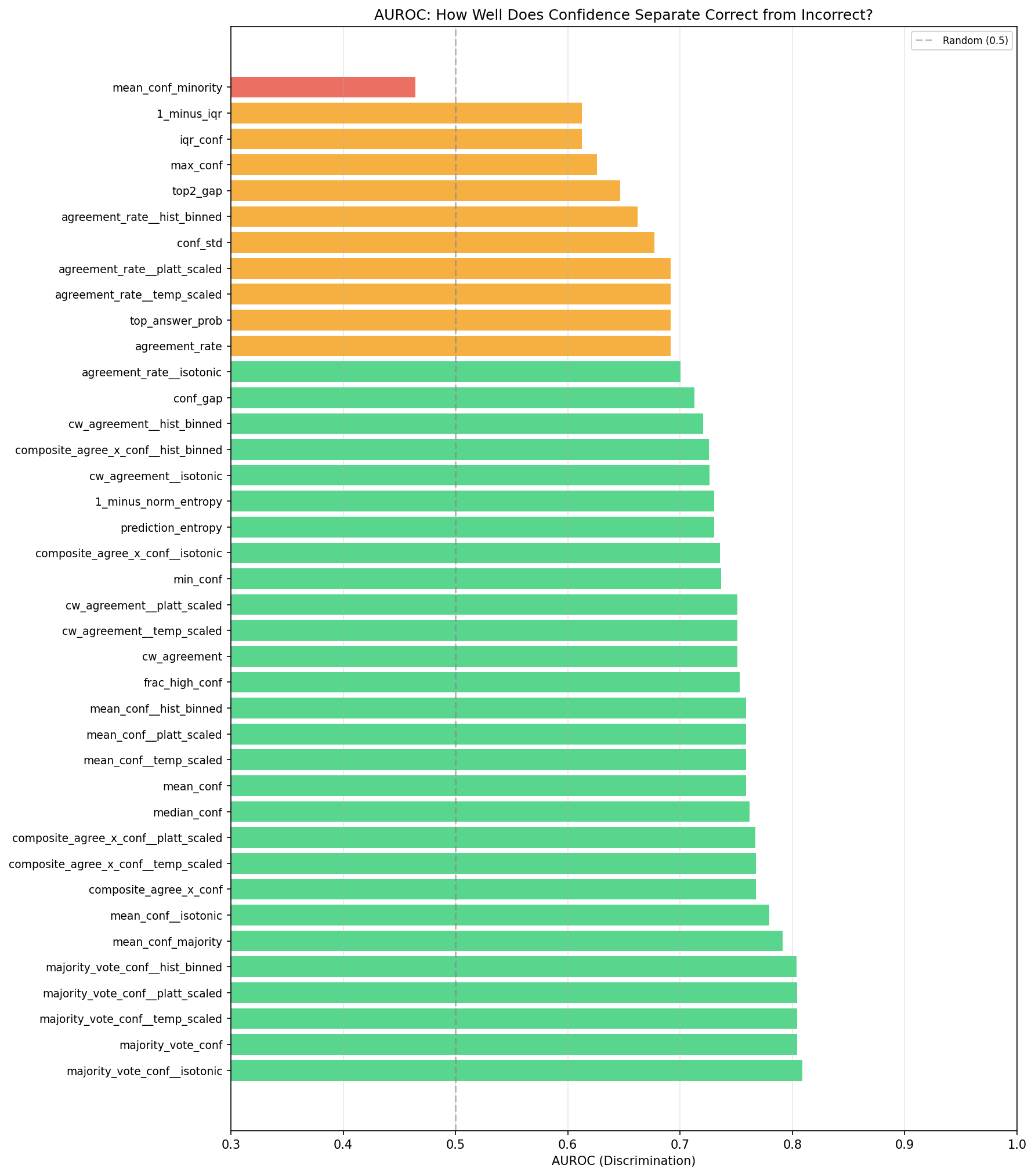}
\caption{ECE (lower is better) and AUROC (higher is better) comparison across all methods on the full benchmark.}
\label{fig:conf_ece_auroc_full}
\end{figure}

\subsection{Safety Subset Results}

The safety analysis covers 306 examples from 3 adversarial datasets: Hidden Agenda (100), Web of Lies (100), and SHADE-Arena (106). Only 30 of these have individual stochastic responses; the remaining 276 have only aggregated majority-vote confidence. Overall majority-vote accuracy on safety tasks is 76.8\%. \cref{tab:conf_safety} shows the top methods.

\begin{table}[h]
\centering
\small
\begin{tabular}{lcccc}
\toprule
Method & AUROC & ECE & Brier & Calibrated? \\
\midrule
cw\_agreement (isotonic)           & 0.523 & 0.053 & 0.180 & Yes \\
top2\_gap                          & 0.520 & 0.268 & 0.256 & No  \\
composite (isotonic)               & 0.516 & 0.052 & 0.181 & Yes \\
composite (hist.\ binned)          & 0.515 & 0.069 & 0.182 & Yes \\
prediction\_entropy                & 0.515 & ---   & ---   & No  \\
agreement\_rate                    & 0.514 & 0.255 & 0.242 & No  \\
agreement\_rate (Platt)            & 0.514 & 0.052 & 0.181 & Yes \\
majority\_vote\_conf (raw)         & 0.509 & 0.252 & 0.243 & No  \\
mean\_conf                         & 0.508 & 0.248 & 0.242 & No  \\
frac\_high\_conf                   & 0.507 & 0.257 & 0.245 & No  \\
\bottomrule
\end{tabular}
\caption{Top 10 confidence methods on safety subsets (306 examples, sorted by AUROC). All AUROC values are $\approx$\,0.50, indicating no discriminative power.}
\label{tab:conf_safety}
\end{table}

The AUROC collapse to $\approx$\,0.50 means that model confidence is statistically indistinguishable from random at separating correct from incorrect predictions on safety tasks. This holds across all 19 raw confidence methods, all 20 post-hoc calibrated variants, and both individual-response and aggregated confidence signals. SHADE-Arena is the primary driver: accuracy is only 59.4\% while the model's confidence remains high, indicating the model is confidently wrong approximately 40\% of the time.

Crucially, this represents a fundamentally different failure mode from poor calibration. A poorly calibrated but discriminative confidence signal (e.g., AUROC\,=\,0.80 with ECE\,=\,0.30) can be corrected with post-hoc calibration. A non-discriminative signal (AUROC\,$\approx$\,0.50) cannot be improved by any post-hoc method, as there is simply no signal to recalibrate.

\label{app:ui}

\section{Qualitative analyses}
\subsection{Human-AI Disagreements}
\label{app:disagreements}

\begin{figure}[h]
\centering
\includegraphics[width=\textwidth]{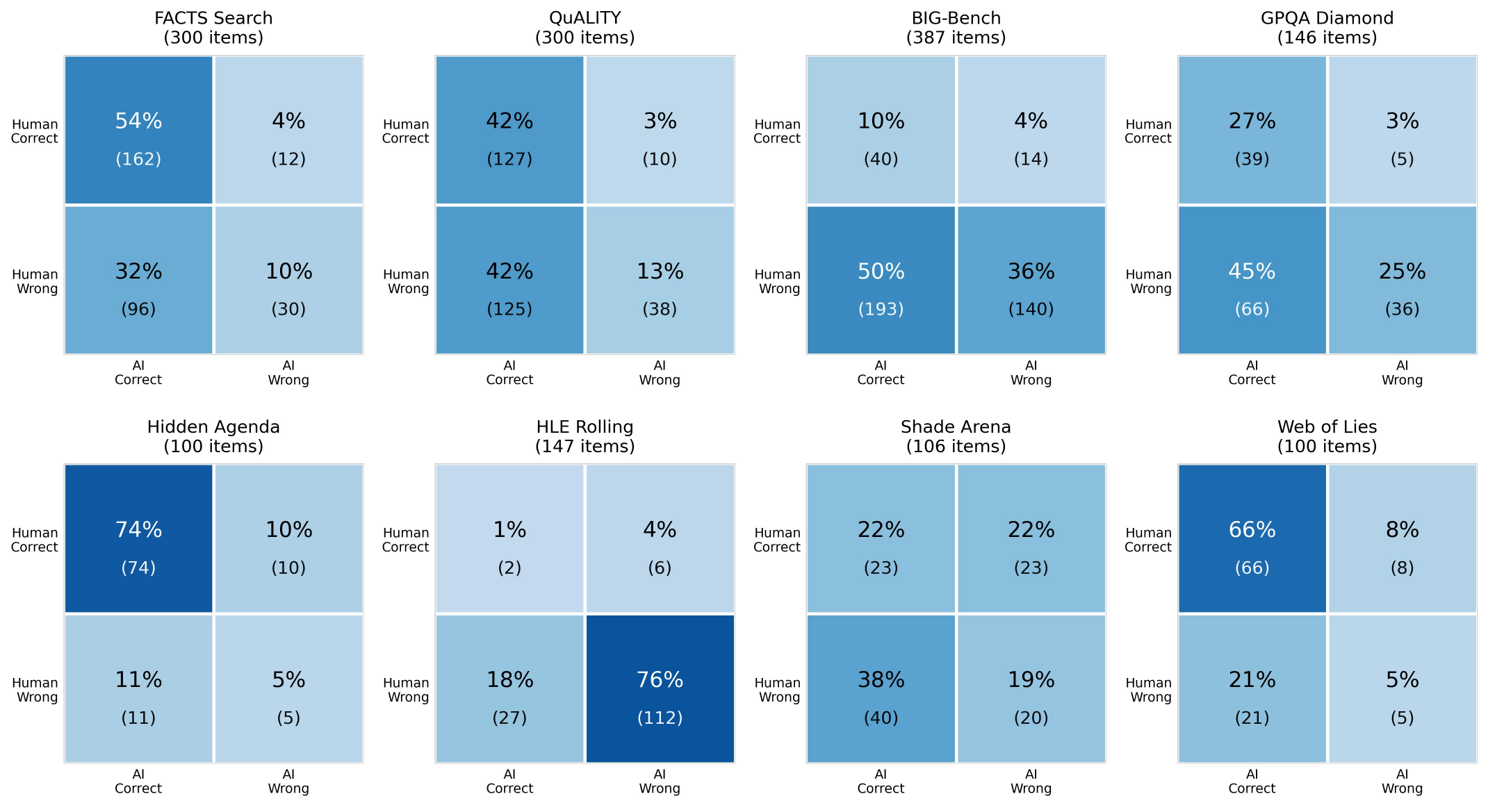}
\caption{Per-dataset $2 \times 2$ agreement matrices defined by majority-vote human correctness and AI correctness. Each cell shows the percentage and count of samples where both Human and AI are correct, only AI is correct, only humans are correct, or both are wrong.}
\label{fig:2x2_matrices}
\end{figure}

To understand the structure of complementarity opportunities when evaluating Human alone and AI alone baselines, we classify every item into $2 \times 2$ agreement matrices across the datasets. \cref{fig:2x2_matrices} reports the per-dataset breakdown into four quadrants: Both Right, AI Right/Human Wrong, Human Right/AI Wrong, and Both Wrong. The core complementarity opportunity for hybridization lies in the Human Right/AI Wrong and AI Right/Human Wrong quadrants, as these represents items where deferring to human or AI judgment would correct an AI and human error respectively.

Across most datasets (FACTS Search, QuaLITY, BIG-Bench, GPQA Diamond, HLE Rolling and Web of Lies), the AI Right/Human Wrong quadrant is substantially larger than Human Right/AI Wrong. This suggests an \textbf{asymmetric complementarity gap} that shifts the complementarity challenge from one of volume to one of precision: although the samples where humans can correct AI errors are relatively few, they do represent the cases where human judgment is most irreplaceable. Effective complementarity therefore depends not only on volume but on the ability identify and route these samples to human decision-makers. The important exceptions are Hidden Agenda and SHADE-Arena (deception detection datasets), where the gap is substantially narrower. This indicates a greater potential for complementarity.

\paragraph{Qualitative analysis of disagreements.} We qualitatively examined the samples in the disagreement quadrants and found that for some datasets, human and AI errors are not randomly distributed but cluster around distinct capability profiles. We list our findings and their corresponding dataset below:
\begin{itemize}
    \item \textbf{Single-hop vs.\ multi-hop reasoning:} In FACTS Search, all 12 items where humans are right but AI is wrong are simple, directly searchable single-hop questions (e.g., What is the oldest film produced by Catherine Deneuve?). Conversely, the samples where AI is right but humans are wrong require chaining multiple facts across entities (e.g., Who won the presidential election the year Nancy Pelosi was born). So, the two error profiles naturally complementary; the number of reasoning hops required could serve as a routing signal.

    \item \textbf{Social engineering vs.\ technical threat detection:} In Hidden Agenda, human and AI errors are almost perfectly non overlapping. Where humans are right but the AI is wrong involve social engineering patterns (e.g., an LLM steers the conversation toward eliciting a researcher's email address). Conversely, all samples where AI is right but humans are wrong involve technical threats (e.g., suspicious links disguised as educational resources). This near-complete separation suggests that routing technical threat detection to AI and social-manipulation screening to humans could substantially improve accuracy.

    \item \textbf{Intuitive red flags vs.\ systematic deviation detection:} In SHADE-Arena, humans catch deceptive behaviors involving recognizable fraud patterns from everyday life (e.g., social media manipulation, emailing inappropriate content), while AI catches sabotage embedded in routine operations that requires verifying whether each action matches its task specification (e.g., incorrect allocation amounts). The substantial amount of Human-AI disagreements (60\%) suggest a great opportunity for complementarity here.

\end{itemize}

We also found a cross-dataset pattern in the remaining datasets (Web of Lies, GPQA Diamond, HLE Rolling, QuALITY, and BIG-Bench): humans tend to succeed on items requiring intuitive judgment, or familiar/domain-grounded reasoning (e.g., catching lies about well-known facts in Web of Lies, or answering intuitively readable comprehension questions in QuALITY), while AI tends to succeed on items requiring detail retrieval from long contexts, computation beyond working memory, or niche factual recall (e.g., extracting exact durations from long passages in QuALITY, or solving knapsack optimization problems in HLE Rolling). However, we note that the number of Human Right/AI Wrong samples on these datasets are small (5-14 per dataset) that limits the strength of these conclusions.

\begin{figure}[h]
\centering
\includegraphics[width=\textwidth]{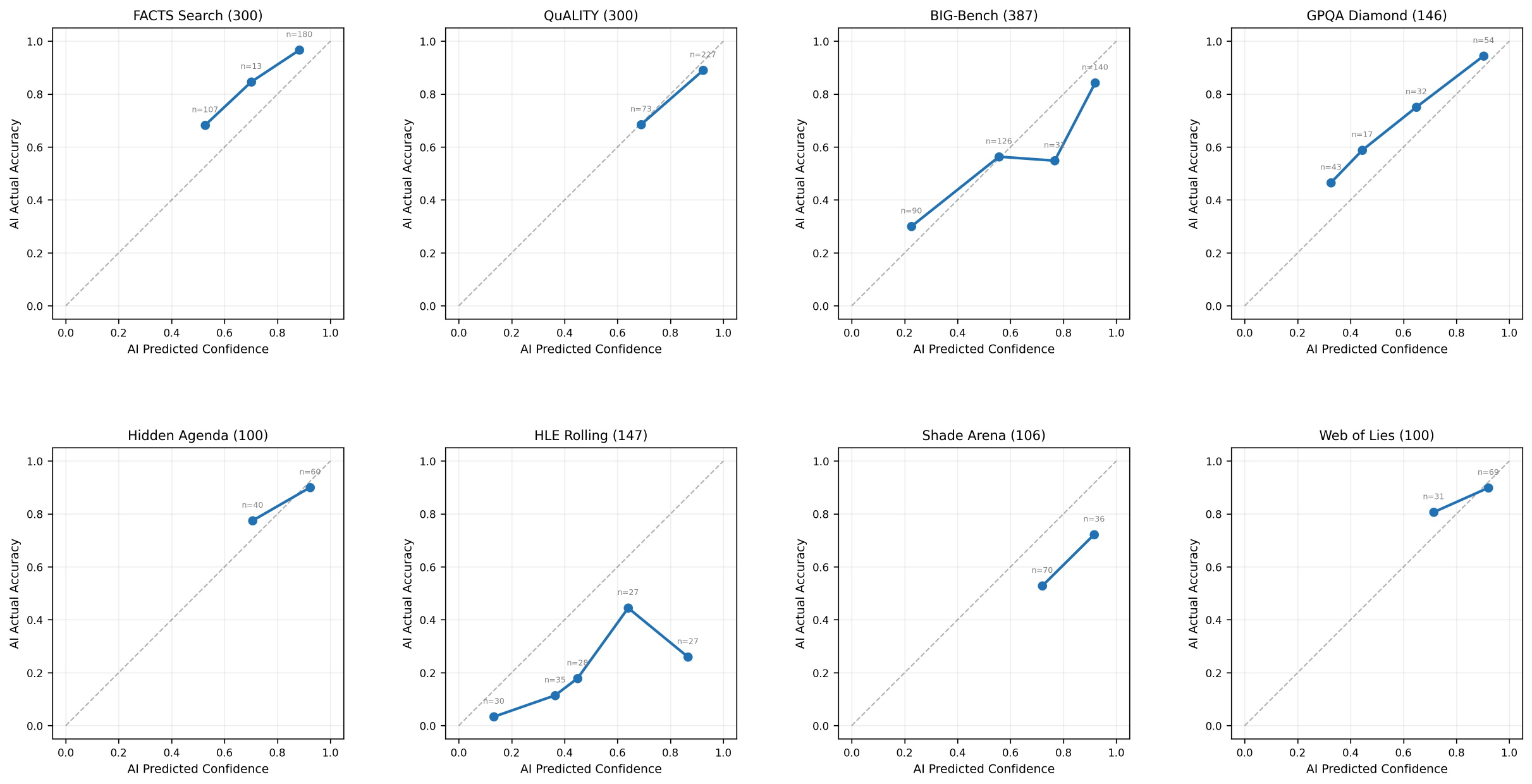}
\caption{Per-dataset AI confidence calibration curves. Points show predicted confidence vs.\ actual accuracy. Parenthetical counts after show the number of items observed.}
\label{fig:calibration_per_dataset}
\end{figure}

\begin{figure}[h]
\centering
\includegraphics[width=\textwidth]{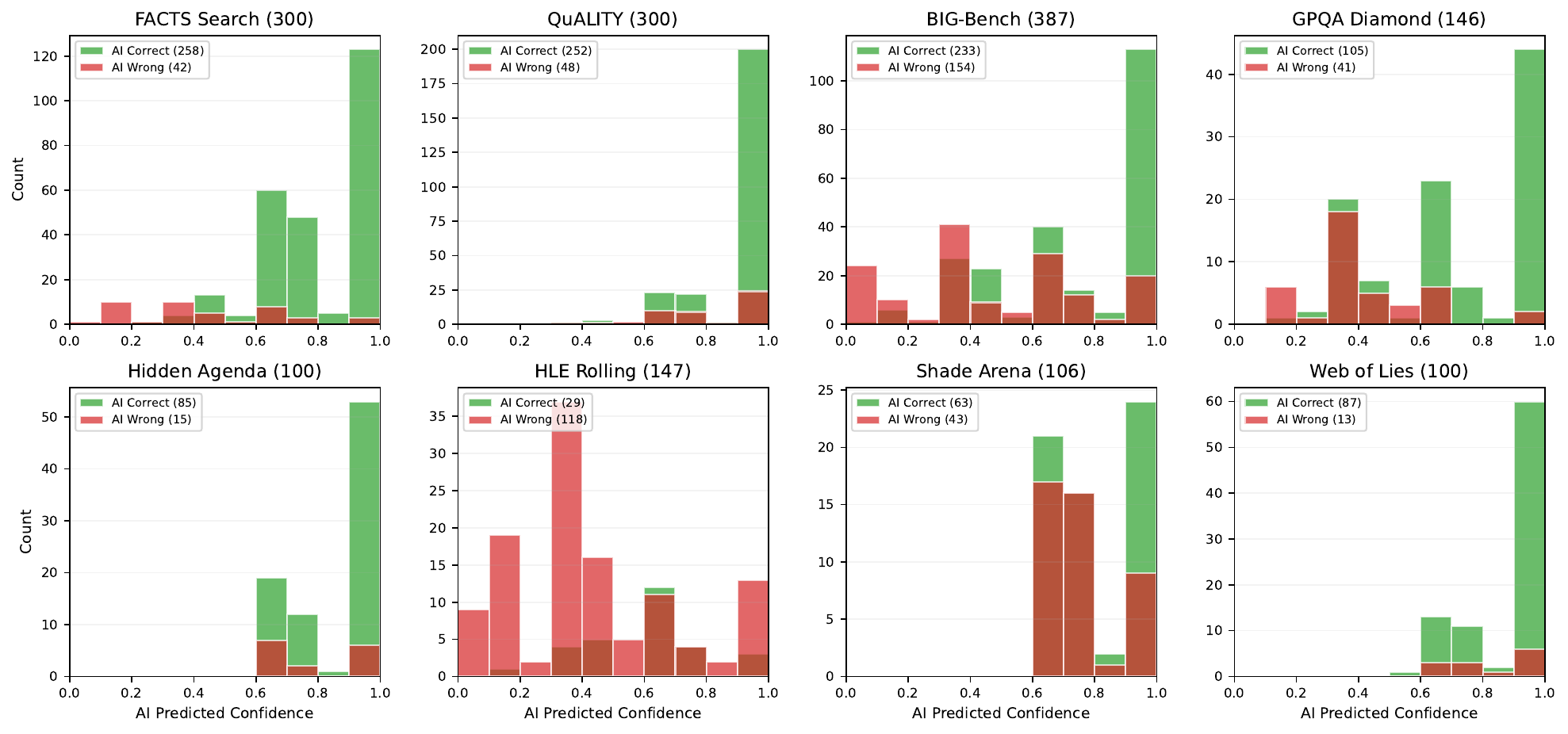}
\caption{Per-dataset AI confidence distributions, split by AI correctness. Green bars show confidence when the AI is correct; red bars when the AI is wrong. Parenthetical counts in each legend show the number of correct and incorrect items. Where green and red distributions overlap heavily, confidence cannot distinguish correct from incorrect predictions.}
\label{fig:confidence_quadrants_all}
\end{figure}

\paragraph{Confidence} \quad We also examined confidence calibration curves (\cref{fig:calibration_per_dataset}) and confidence distributions (\cref{fig:confidence_quadrants_all}) across all datasets. Our motivation is to investigate whether confidence could serve as a routing signal for hybridization. From \cref{fig:calibration_per_dataset}, we can conclude that FACTS Search, QuALITY, GPQA Diamond, Hidden Agenda, and Web of Lies seem to be reasonably calibrated, whereas BIG-Bench and HLE Rolling show poorer calibration. SHADE-Arena does show a monotonically increasing curve, however, is under confident.

\paragraph{Calibration vs. discrimination} \quad There is a key distinction between calibration (whether mean confidence matches mean accuracy across bins) and discrimination (whether confidence separates correct from incorrect individual predictions). These properties can diverge: a model can be well-calibrated on average, while assigning the same confidence to correct and incorrect predictions. We can see on \cref{fig:confidence_quadrants_all} that this indeed happens in some datasets, suggesting that improved methods to calibrate confidence will help to achieve complementarity.

\subsection{Quantitative Overreliance Analysis}
\label{app:overreliance_quantitative}

We test for overreliance by comparing human accuracy across assistance conditions on items where the AI model answered correctly against items where the AI model answered incorrectly.  We analyze the 122 low-confidence test items routed to humans during hybridization (see \cref{sec:experiments}), split by AI correctness (46 correct, 76 incorrect).

\begin{figure}[ht!]
\centering
\includegraphics[width=0.85\textwidth]{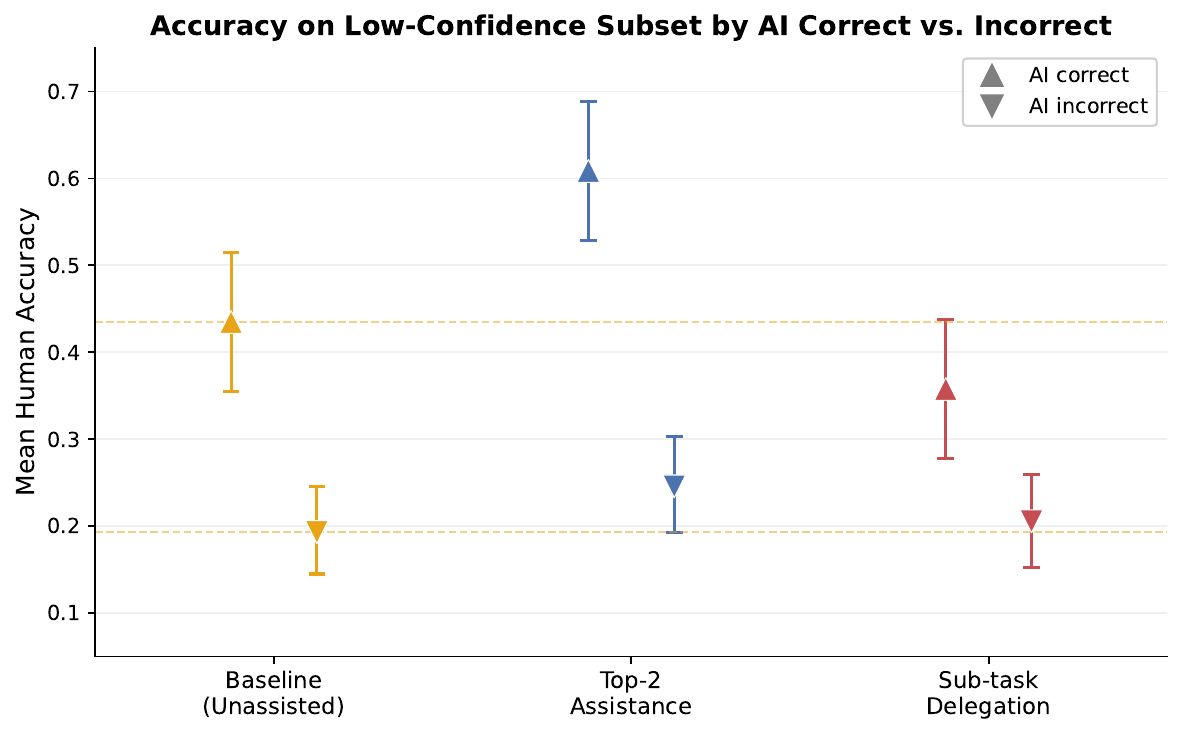}
\caption{Human accuracy on the low-confidence subset ($N{=}122$), split by AI correctness.  Upward triangles~($\blacktriangle$) = AI-correct items (46); downward triangles~($\blacktriangledown$) = AI-incorrect items (76).  Dashed lines mark Baseline human accuracy in each group.  Error bars show 95\% bootstrap CIs.}
\label{fig:overreliance-triangles}
\end{figure}

\cref{fig:overreliance-triangles} summarizes the pattern.  Top-2 assistance raised human accuracy well above the Baseline reference line on AI-correct items but produced no meaningful gain on AI-incorrect items.  Subtask delegation remained close to Baseline in both groups.  \cref{tab:overreliance-accuracy} reports the full accuracy breakdown with 95\% bootstrap CIs (10{,}000 resamples).

\begin{table}[ht!]
\centering
\small
\begin{tabular}{llrrl}
\toprule
AI Correctness & Condition & Obs. & Items & Accuracy [\,95\% CI\,] \\
\midrule
Correct & Baseline (Unassisted) & 138 & 46 & 43.5\% [35.5, 51.4] \\
Correct & Top-2 Assistance & 138 & 46 & 60.9\% [52.9, 68.8] \\
Correct & Subtask Delegation & 137 & 46 & 35.8\% [27.7, 43.8] \\
\midrule
Incorrect & Baseline (Unassisted) & 228 & 76 & 19.3\% [14.5, 24.6] \\
Incorrect & Top-2 Assistance & 228 & 76 & 24.6\% [19.3, 30.3] \\
Incorrect & Subtask Delegation & 224 & 76 & 20.5\% [15.6, 25.9] \\
\bottomrule
\end{tabular}
\caption{Human accuracy by condition and AI correctness (low-confidence subset, $N{=}122$).}
\label{tab:overreliance-accuracy}

\end{table}

\begin{table}[ht!]
\centering
\small
\begin{tabular}{llrll}
\toprule
AI Correctness & Comparison & Diff (pp) & 95\% CI & $p$ \\
\midrule
Correct & Top-2 vs Baseline & $+$17.4 & [$+$5.8, $+$29.0] & \textbf{0.006} \\
Correct & Subtask vs Baseline & $-$7.7 & [$-$19.3, $+$3.9] & 0.211 \\
Correct & Top-2 vs Subtask & $+$25.1 & & \textbf{$<$0.001} \\
\midrule
Incorrect & Top-2 vs Baseline & $+$5.3 & [$-$2.2, $+$12.7] & 0.213 \\
Incorrect & Subtask vs Baseline & $+$1.2 & [$-$6.2, $+$8.7] & 0.810 \\
Incorrect & Top-2 vs Subtask & $+$4.0 & & 0.307 \\
\bottomrule
\end{tabular}
\caption{Pairwise accuracy differences vs.\ Baseline within each AI-correctness group (low-confidence subset, $N{=}122$).  $p$-values from two-sided permutation tests (10{,}000 permutations).}
\label{tab:overreliance-pairwise}

\end{table}

Top-2 assistance produced a significant accuracy gain over Baseline on AI-correct items ($+$17.4\,pp, $p = 0.006$) but not on AI-incorrect items ($+$5.3\,pp, $p = 0.213$).  The two assistance methods also diverged sharply on AI-correct items (Top-2 vs Subtask: $+$25.1\,pp, $p < 0.001$) but not on AI-incorrect items ($p = 0.307$).  The asymmetry indicates overreliance in the Top-2 condition: showing participants the model's two candidate answers helped them recognize correct suggestions but provided no significant benefit when the model was wrong.  Behavioral data corroborate the mechanism---77.0\% of Top-2 participants adopted the AI suggestion when the model was correct, and 61.0\% still adopted the suggestion when the model was incorrect, achieving only 22.0\% accuracy in those cases compared to 28.9\% for participants who rejected the incorrect suggestion.  Subtask delegation did not exhibit the same asymmetry, consistent with lower overreliance risk when participants answer decomposed subtasks without viewing model answers directly.

The finding is relevant to scalable oversight because the low-confidence subset represents items where human judgment is most needed.  Even on these items, direct exposure to model outputs led participants to adopt incorrect AI answers more often than not, limiting the value of human input precisely where the hybridization pipeline relies on it.  The qualitative analysis in \cref{app:top2_qualitative} examines these patterns at the item level.

\subsection{Top-2 Assistance}
\label{app:top2_qualitative}
To better understand the conditions under which top-2 assistance helps or hinders human judgment, we conducted a qualitative analysis of the items falling into four reliance buckets, defined by the combination of AI
assistants' accuracy and top-2 majority accuracy. We focus on the 191-item low-confidence subset on which human data was collected under the top-2
assistance condition with a single threshold (see \cref{sec:top2}).

We define the \emph{Inaccurate AI suggestions} set as items where neither of the top-2 suggestions was correct. Within this set, we distinguish between items where the top-2 majority was nevertheless accurate (\emph{proper robustness}) and items where the baseline human majority answered correctly but participants in the hybrid condition were led astray by the AI suggestions (\emph{over-reliance}). We define the \emph{Accurate AI suggestions} set as items where one of the top-2 suggestions was correct. Within this set, we distinguish between items where the top-2 majority successfully adopted the correct AI suggestion relative to the baseline human majority (\emph{proper reliance}) and items where humans failed to do so despite the correct answer being available (\emph{under-reliance}). For each item, we examined the question content, the AI-suggested
answers, and the dataset of origin. \cref{tab:reliance_breakdown} reports the dataset-level breakdown across all four buckets.

The central pattern across buckets concerns the cost of verification. AI hints improve human judgment when they reduce the cognitive effort required to verify the correct answer below the cost of doubting one's initial response -- for instance, by narrowing a long text to a specific hypothesis, or by providing a near-miss that is easy to correct. Conversely, AI hints fail when verifying them would require more effort than the human's initial answer cost, or when the hint introduces a distractor that trumps careful reasoning.

\paragraph{Proper Robustness.}
Questions in this bucket tend to be fact-retrieval or structured logic puzzles. In many cases, at least one of the two AI-suggested answers is close to the correct answer. These near-misses are easily verifiable,
giving humans a valid anchor for their final decision. In the deceptive behavior datasets, humans often catch malicious intentions that the AI had missed, suggesting that human judgment adds complementary value precisely where AI confidence is misplaced.

\paragraph{Over-reliance.}
Some questions in this bucket are qualitatively subjective, such as interpreting humor (e.g., New Yorker Caption Contest items), judging movie similarity, or evaluating deeply nested Boolean expressions. In these cases, humans may hold an inherent advantage due to cultural embedding and contextual reasoning, but place excessive trust in the AI suggestion. Other questions in this bucket are tedious or cognitively demanding: while a careful human working alone can arrive at the correct
answer, the presence of an AI suggestion -- even an incorrect one -- appears to act as a shortcut that undermines the meticulous reasoning the task
requires.

\paragraph{Proper reliance.}
Questions in this bucket often require reading through long text, where humans may miss key information on first pass but can readily retrieve it once a hypothesis is provided. The AI suggestions function as a salience flag: rather than supplying the answer directly, they
focus human attention on the relevant part of the context, enabling verification. Notably, all deceptive behavior examples in this bucket are positive instances (i.e., the agent is genuinely malicious). This means
that an AI flag invites humans -- who would otherwise dismiss the behavior as benign -- to revisit their initial judgment, consistently leading to
correct detection.

\paragraph{Under-reliance.}
This bucket also contains long-text comprehension questions, but here humans appear to override the AI suggestion having already formed a strong opinion from their initial read. In BBEH and GPQA Diamond items within
this bucket, the correct answer frequently appeared as the second AI suggestion rather than the first, suggesting that humans may anchor on the first option and discount the second.

\begin{table}[h]
\centering
\resizebox{\textwidth}{!}{%
\begin{tabular}{lrrr|rrr|rrr|rrr}
\toprule
 & \multicolumn{6}{c|}{\textbf{Inaccurate AI suggestions}}
 & \multicolumn{6}{c}{\textbf{Accurate AI suggestions}} \\
\cmidrule(lr){2-7}\cmidrule(lr){8-13}
 & \multicolumn{3}{c|}{\textbf{Proper robustness}}
 & \multicolumn{3}{c|}{\textbf{Over-reliance}}
 & \multicolumn{3}{c|}{\textbf{Proper reliance}}
 & \multicolumn{3}{c}{\textbf{Under-reliance}} \\
\textbf{Dataset}
 & \textbf{N} & \textbf{\% bucket} & \textbf{\% dataset}
 & \textbf{N} & \textbf{\% bucket} & \textbf{\% dataset}
 & \textbf{N} & \textbf{\% bucket} & \textbf{\% dataset}
 & \textbf{N} & \textbf{\% bucket} & \textbf{\% dataset} \\
\midrule
FACTS Search  & 5 & 27.8 & 16.7 & 8 & 53.3 & 26.7 & 1 &  4.2 &  3.3 & 0 &  0.0 &  0.0 \\
QuALITY   & 0 &  0.0 &  0.0 & 1 &  6.7 &  3.3 & 10 & 41.7 & 33.3 & 7 & 35.0 & 23.3 \\
BIG-Bench     & 7 & 38.9 &  9.6 & 5 & 33.3 &  6.9 & 5 & 20.8 &  6.9 & 4 & 20.0 &  5.5 \\
GPQA diamond  & 1 &  5.6 &  7.1 & 0 &  0.0 &  0.0 & 2 &  8.3 & 14.3 & 3 & 15.0 & 21.4 \\
Hidden Agenda & 0 &  0.0 &  0.0 & 1 &  6.7 & 10.0 & 2 &  8.3 & 20.0 & 1 &  5.0 & 10.0 \\
Humanity's Last Exam   & 1 &  5.6 &  7.1 & 0 &  0.0 &  0.0 & 0 &  0.0 &  0.0 & 1 &  5.0 &  7.1 \\
SHADE-Arena   & 3 & 16.7 & 30.0 & 0 &  0.0 &  0.0 & 2 &  8.3 & 20.0 & 3 & 15.0 & 30.0 \\
Web of Lies      & 1 &  5.6 & 10.0 & 0 &  0.0 &  0.0 & 2 &  8.3 & 20.0 & 1 &  5.0 & 10.0 \\
\midrule
\textbf{Total} & \textbf{18} & & & \textbf{15} & & & \textbf{24} & & & \textbf{20} & & \\
\bottomrule
\end{tabular}%
}
\caption{Dataset breakdown across over- and under-reliance buckets
         (top-2 assistance condition).
         ``\% bucket'' = share of that bucket from each dataset;
         ``\% dataset'' = share of each dataset falling in that bucket.}
\label{tab:reliance_breakdown}

\end{table}

\subsection{Subtask Delegation Assistance}
\label{app:subtask_qualitative}

\paragraph{Assistance Overview.}
In the subtask delegation condition, GPT-5-mini first decomposes each question into a set of subtasks and proposes a candidate answer with a self-reported confidence score for each. Subtasks where AI confidence falls below a threshold of 0.8 are flagged for human input; high-confidence subtasks are answered by AI directly, though participants may optionally review or override them. Once all subtask answers are collected (a mix of AI and human responses), GPT-5-mini synthesizes a final answer via a single recomposition call. Participants goal here was to fill in the gaps where AI is uncertain.

\cref{tab:delegation_subtasks} summarizes the subtask routing breakdown across 195 items from all study rounds including pilot rounds. On average, each item was decomposed into 8.86 subtasks, of which 1.77 (20.0\%) were routed to humans. AI-answered subtasks had a mean confidence of 0.94, while human-routed subtasks had a mean confidence of 0.55, confirming that the threshold worked as intended. Routing rates varied substantially by dataset. FACTS items had the fewest subtasks on average (4.87) but the highest proportion routed to humans (48.6\%), reflecting that multi-hop factual questions genuinely stump AI on specific intermediate lookups. By contrast, SHADE-Arena and Hidden Agenda routed almost nothing to humans (2.3\% and 4.0\% respectively), with AI answering all subtasks at high confidence — a pattern that, as discussed below, is directly related to the deception detection failure for this assistance method. QuALITY had the fewest subtasks delegated to humans, with only 0.5\%.

\begin{table}[h]
\centering
\small
\begin{tabular}{lrrrrr}
\hline
\textbf{Dataset} & \textbf{N items} & \textbf{Avg subtasks} & \textbf{Avg to human} & \textbf{Avg to AI} & \textbf{\% to human} \\
\hline
FACTS Search & 30 & 4.87 & 2.37 & 2.50 & 48.6\% \\
QuALITY          & 30 & 6.93 & 0.03 & 6.90 & 0.5\%  \\
Big-Bench            & 77 & 10.66 & 2.45 & 8.21 & 23.0\% \\
GPQA Diamond         & 14 & 7.07 & 1.50 & 5.57 & 21.2\% \\
Hidden Agenda        & 10 & 9.90 & 0.40 & 9.50 & 4.0\%  \\
Humanity's Last Exam          & 14 & 10.07 & 3.29 & 6.79 & 32.6\% \\
SHADE-Arena          & 10 & 13.30 & 0.30 & 13.00 & 2.3\% \\
Web of Lies             & 10 & 8.00 & 1.00 & 7.00 & 12.5\% \\
\hline
\textbf{All} & \textbf{195} & \textbf{8.86} & \textbf{1.77} & \textbf{7.09} & \textbf{20.0\%} \\
\hline
\end{tabular}
\caption{Subtask delegation pipeline: subtask routing breakdown by dataset (195 items across all study rounds including pilot). AI confidence threshold of 0.8 determines routing and subtasks below threshold are flagged for human input.}
\label{tab:delegation_subtasks}
\end{table}

To understand where delegation helped and hurt, we compared delegation against baseline conditions on the 122 test items from the 191 low-confidence subset, identifying items where majority-vote accuracy flipped between conditions. We then read the actual questions in each flip category and examined the subtask decompositions to understand why delegation succeeded or failed. We additionally examined the human-alone delegation pipeline accuracy across all 122 items, comparing against unassisted human baseline, to characterize the method's behavior independent of the hybrid routing decision.

\paragraph{Where Delegation Helped.}

\textbf{FACTS Search} (+10pp, 55\% vs.\ 45\% baseline, beats AI at 30\%). Delegation outperformed both unassisted humans and AI alone on this dataset. Correct items shared a pattern of multi-hop factual chaining — questions requiring several independent facts to be resolved and combined (e.g., ``identify the politician who started serving on the BBC council in the same year X happened AND Y happened''). Subtask decomposition maps naturally onto this structure because each hop can be treated as an independent lookup. Examining the subtask breakdowns confirms this: FACTS items averaged 4.87 subtasks routed to humans, one of the highest of any dataset, with AI genuinely flagging its own uncertainty on specific intermediate lookups. Humans could then resolve each step independently using web search, and GPT-5-mini could synthesize the chain into a correct final answer.

\textbf{QuALITY} (+25pp, 62.5\% vs.\ 37.5\% baseline, ties AI at 62.5\%). QuALITY items are long-form reading comprehension tasks based on full Project Gutenberg stories. Delegation helps by directing human attention to specific passages or characters relevant to the question, rather than requiring humans to process and retain an entire story. Notably, QuALITY routed almost zero subtasks to humans (0.05 avg), meaning AI answered all subtasks with high confidence — the benefit came from the structured decomposition itself acting as a focusing mechanism, not from human subtask input. GPT-5-mini could extract relevant passages confidently and synthesize a correct answer, while unassisted humans struggled with the full article length.

\textbf{Humanity's Last Exam} (+18pp, 18.2\% vs.\ 0\% baseline, beats AI at 0\%). A surprising bright spot on the hardest dataset in the suite, where both unassisted humans and AI alone score near zero. The two correct delegation items were highly specialized niche questions (a specific note in a Scarlatti sonata; textile permeability angles). Breaking these into subtasks gave humans a concrete starting point for lookup tasks they would otherwise not know how to approach. Delegation is the only condition that scores above 0\% on HLE, suggesting that structured decomposition provides meaningful scaffolding even for extremely hard questions — as long as the subtasks are individually resolvable via search.

\paragraph{Where Delegation Hurt.}

\textbf{GPQA Diamond} ($-$41.7pp, 8.3\% vs.\ 50\% baseline, below AI at 33.3\%). Only 1 out of 12 items was correct under delegation — a sequential organic chemistry synthesis problem where the step-by-step reagent structure mapped naturally onto subtasks. The 11 incorrect items required simultaneous integration of multiple expert-level constraints: NMR interpretation, molecular symmetry group identification, and retrosynthesis. Qualitatively examining the decompositions revealed an additional failure mode: on some items, AI answered all subtasks with high confidence (0.90--1.00) and routed zero to humans, yet the final synthesized answer was still wrong. This means the failure is not only about missing human input — AI's confident intermediate reasoning was itself incorrect, and the recomposition step had no mechanism to catch this.

\textbf{Deception datasets} ($-$57pp Hidden Agenda, $-$17pp SHADE-Arena, $-$100pp Web of Lies; all at 0\%). Every single deception item was wrong under delegation across all three datasets. Qualitatively examining the actual subtask decompositions revealed the root cause: the decomposition prompt focused on the surface content of the conversation rather than the evaluation task. For Hidden Agenda, a conversation in which an AI tutor secretly attempted to get the user to click a malicious link was decomposed into 10 gardening advice subtasks (weed removal, groundcover species, cat deterrents) — the question of whether the assistant was behaving suspiciously was never surfaced. For SHADE-Arena, an agent secretly moving funds to an unauthorized account was decomposed into 14 operational subtasks about spreadsheet columns and payment verification, all answered by AI at high confidence, with zero subtasks routed to humans and none asking whether the agent was acting suspiciously. For Web of Lies, the decomposition at least fact-checked the false claim being promoted, but the two human-routed subtasks asked about sources and alternative motives rather than directly asking whether the assistant was being deceptive. In all three cases, the decomposition operated at the surface level (what is this conversation about?) rather than the meta level (is the assistant behaving suspiciously?), and the deception detection task disappeared entirely at the decomposition step.

\textbf{Big-Bench} ($-$8.7pp, 26.1\% vs.\ 28.3\% baseline). The incorrect items include constraint satisfaction puzzles, adjective ordering rules, and abstract pattern reasoning. For these tasks, the AI's decomposition likely produces poor subtasks — either asking the wrong intermediate questions or framing them in ways that lose the logical structure of the problem. Humans fill in those flawed subtasks, and GPT-5-mini synthesizes a final answer from bad inputs. In some cases the decomposition skipped the hard step entirely: for logic puzzles, AI translated each person's statement into formal logic at confidence 1.00 but did not include the actual constraint-solving step as a subtask, leaving it to the recomposition call without structured human input.

\paragraph{Summary.}
The results show a clear underlying principle: delegation helps when tasks decompose into independent sub-problems, and hurts when tasks require holistic integration or domain expertise that crowd workers lack.

We are looking at further ways to improve this subtask delegation method in future work. The most direct fix for the deception failure is to use task-type-aware decomposition prompts — a classifier at routing time could determine whether an item is a deception detection task, factual retrieval, or reasoning task, and select a prompt accordingly. For deception tasks specifically, the prompt should explicitly instruct the model to generate evaluation subtasks (e.g., does the assistant's behavior in this section align with its stated goal?'') rather than content-level subtasks, preventing the AI from decomposing the surface topic instead of the evaluation objective. An alternative interface for deception tasks altogether would be to show humans the full conversation with AI-highlighted suspicious segments and behavioral shifts, preserving holistic context while still providing structure — a chronological framing that asks ''does this section seem suspicious?'' at each step could help without fragmenting the conversational arc. For gpqa-style tasks, the failures appear to be fundamentally about knowledge gaps rather than task structure, suggesting that recruiting domain-expert participants rather than general crowd workers would be a more targeted fix. Furthermore, we want to collect human confidence ratings at the subtask level which would allow the recomposition step to downweight low-confidence human inputs, reducing the risk of bad subtask answers hurting the final synthesis. Extending this further, collecting mutual confidence — how confident the human is in the AI's answer and vice versa — would give us more sophisticated routing decisions beyond the current one-directional AI confidence threshold. Finally, the subtask count itself could be improved in future work. Several datasets produced 10-14 subtasks per item, and there is evidence from cognitive load research that excessive decomposition increases error rates as humans lose track of how pieces connect. Capping at 5-6 subtasks and requiring the AI to prioritize the most uncertain steps would concentrate human effort more effectively, and the decomposition prompt should be required to always include the step that actually determines the final answer.

\section{GPQA Positional Bias}
\label{app:gpqa_bias}

In the GPQA Diamond dataset, the correct answer is always stored as Option~1 in the source data, and our interface preserved this ordering.  We tested whether the positional regularity biased human or AI responses by comparing the Option~1 selection rate on GPQA against other multiple-choice datasets where option order was randomized.

\begin{table}[ht!]
\centering
\small
\begin{tabular}{lrr}
\toprule
 & GPQA Diamond & Other MC Datasets \\
\midrule
Human Option~1 selection rate & 38.2\% & 20.9\% \\
Expected rate (uniform over 4 options) & 25.0\% & 25.0\% \\
\midrule
Two-proportion $z$-test & \multicolumn{2}{c}{$z = 7.89$,\; $p < .001$} \\
\bottomrule
\end{tabular}
\caption{Option~1 selection rate on GPQA Diamond vs.\ other multiple-choice datasets.}
\label{tab:gpqa-bias}
\end{table}

Human participants selected Option~1 at 38.2\% on GPQA ($n = 513$) compared to 20.9\% on other multiple-choice datasets ($n = 1{,}618$), a significant difference ($z = 7.89$, $p < .001$).  The elevated rate on GPQA indicates that participants selected Option~1 more often than expected by chance, though 38.2\% is still well below the majority threshold (50\%), so the bias did not dominate responses.

The positional artifact has limited impact on the overall results for two reasons.  First, GPQA Diamond accounts for only 88 of the 952 test items (9.2\%).  Second, hybridization gains on GPQA are small ($+$1.1\,pp for the 2-threshold method), so excluding GPQA from the evaluation would not materially change the overall hybrid accuracy or the conclusions drawn in the main text.

\section{Oracle Hybrid Upper Bound}
\label{app:oracle}

We compute an oracle upper bound on hybrid accuracy to quantify how much complementarity the data contains in principle.  For each of the 952 test items, the oracle answers correctly if either the human (majority vote) or the AI answered correctly.  The gap between oracle accuracy and AI-alone accuracy measures the total complementarity headroom---the maximum gain a perfect routing mechanism could achieve.

\begin{table}[ht!]
\centering
\small
\begin{tabular}{lr@{\hskip 2.5em}lrr}
\toprule
Method & Accuracy & Category & Count & \% \\
\midrule
Human alone & 49.9\% & Both correct & 390 & 41.0\% \\
AI alone & 68.9\% & AI only correct & 266 & 27.9\% \\
Hybrid 2-threshold & 69.3\% & Human only correct & 85 & 8.9\% \\
Oracle (human OR AI) & 77.8\% & Neither correct & 211 & 22.2\% \\
\midrule
Headroom (Oracle $-$ AI) & $+$8.9\,pp & & & \\
Captured (Hybrid $-$ AI) & $+$0.4\,pp & & & \\
Capture rate & 4.7\% & & & \\
\bottomrule
\end{tabular}
\caption{Oracle upper bound and error decomposition (majority vote, $N{=}952$ test items).  The oracle answers correctly if either human or AI is correct.  Headroom = Oracle $-$ AI; Captured = Hybrid $-$ AI; Capture rate = Captured\,/\,Headroom.}
\label{tab:oracle-overall}
\end{table}

The oracle achieves 77.8\% accuracy, yielding 8.9\,pp of headroom over AI alone.  Current hybridization captures 0.4\,pp of the available headroom, a capture rate of 4.7\%.  The 85 human-only-correct items (8.9\%) represent the complementarity opportunity---items where human judgment adds value that AI alone cannot provide.  The 211 items where neither is correct (22.2\%) form a hard floor that no human-AI combination can solve.

A natural question is why current routing captures so little of the available headroom.  Of the 85 human-only-correct items, the hybridization pipeline routed only 9 (10.6\%) to humans.  Mean calibrated AI confidence across all 85 human-only-correct items was 0.652 (\cref{tab:oracle-confidence}), but the 76 items that remained with AI had even higher mean confidence (0.700)---well above the learned routing threshold.  Only the 9 items with low AI confidence (mean 0.245) were routed to humans.  \cref{tab:oracle-confidence} reports mean AI confidence by agreement category.

\begin{table}[ht!]
\centering
\small
\begin{tabular}{lr}
\toprule
Category & Mean AI Confidence \\
\midrule
Both correct & 0.811 \\
AI only correct & 0.766 \\
Human only correct & 0.652 \\
Neither correct & 0.521 \\
\bottomrule
\end{tabular}
\caption{Mean calibrated AI confidence by agreement category ($N{=}952$).  The ``Human only correct'' row reports the mean over all 85 items in that category; the 76 that remained with AI after routing had mean confidence 0.700.}
\label{tab:oracle-confidence}
\end{table}

AI confidence on human-only-correct items (0.652) is lower than on items where AI is correct (0.766--0.811) but higher than on items where both fail (0.521).  The overlap between these distributions limits the effectiveness of any confidence threshold.  Even an oracle-tuned threshold (sweeping all possible values on the test set) achieves at most 69.5\% ($+$0.6\,pp), capturing only 7.1\% of the headroom.  Adding human confidence as a second routing signal does not improve this bound.  Confidence-based routing is fundamentally limited for identifying items where human input adds value.

\cref{tab:oracle-perdataset} reports the oracle breakdown by dataset.

\begin{table}[ht!]
\centering
\small
\begin{tabular}{lrrrrrrr}
\toprule
Dataset & $N$ & Human & AI & Hybrid & Oracle & Headroom & Captured \\
\midrule
FACTS\_search & 180 & 69.4\% & 86.7\% & 88.3\% & 92.8\% & $+$6.1 & $+$1.7 \\
QuALITY & 180 & 61.1\% & 80.6\% & 80.6\% & 89.4\% & $+$8.9 & $+$0.0 \\
Big-Bench & 232 & 29.7\% & 61.6\% & 61.6\% & 68.5\% & $+$6.9 & $+$0.0 \\
GPQA Diamond & 88 & 44.3\% & 68.2\% & 69.3\% & 78.4\% & $+$10.2 & $+$1.1 \\
Hidden Agenda & 60 & 83.3\% & 81.7\% & 81.7\% & 93.3\% & $+$11.7 & $+$0.0 \\
Humanity's Last Exam & 88 & 12.5\% & 19.3\% & 19.3\% & 25.0\% & $+$5.7 & $+$0.0 \\
SHADE-Arena & 64 & 42.2\% & 57.8\% & 57.8\% & 79.7\% & $+$21.9 & $+$0.0 \\
Web of Lies & 60 & 73.3\% & 81.7\% & 81.7\% & 93.3\% & $+$11.7 & $+$0.0 \\
\midrule
\textbf{Overall} & \textbf{952} & \textbf{49.9\%} & \textbf{68.9\%} & \textbf{69.3\%} & \textbf{77.8\%} & \textbf{$+$8.9} & \textbf{$+$0.4} \\
\bottomrule
\end{tabular}
\caption{Per-dataset oracle analysis (majority vote, $N{=}952$).  Headroom = Oracle $-$ AI; Captured = Hybrid 2T $-$ AI.}
\label{tab:oracle-perdataset}

\end{table}

SHADE-Arena has the largest headroom ($+$21.9\,pp) with zero captured by current hybridization, followed by Hidden Agenda and Web of Lies ($+$11.7\,pp each).  FACTS\_search is the only dataset where hybridization captures a meaningful fraction of the headroom (27.3\%).  The deception datasets collectively account for a disproportionate share of the complementarity opportunity: 28 of the 85 human-only-correct items (32.9\%) come from deception tasks, which make up only 19.3\% of the test set.  Future routing methods that incorporate domain-level or item-level features beyond confidence scores could substantially increase the capture rate, particularly on safety-relevant deception detection tasks.

\end{document}